\documentclass[usenatbib]{mn2e}
\usepackage{graphicx}
\def\lsim{\vcenter{\hbox{$<$}\offinterlineskip\hbox{$\sim$}}}

\title[Massive YSOs in the Large Magellanic Cloud]
{Massive Young Stellar Objects in the Large Magellanic Cloud: water masers and 
ESO-VLT 3$-$4\,$\mu$m spectroscopy}

\author[Oliveira, van Loon, Stanimirovi\'c \& Zijlstra]{J.M. Oliveira$^{1}$\thanks{E-mail:
joana@astro.keele.ac.uk}, J.Th. van Loon$^{1}$, S. Stanimirovi\'c$^{2}$ and 
A.A. Zijlstra$^{3}$\\
$^{1}$School of Physical \& Geographical Sciences, 
      Lennard-Jones Laboratories, Keele University, Staffordshire ST5 5BG, UK\\
$^{2}$Radio Astronomy Lab, University of California at Berkeley, 601 Campbell
      Hall, Berkeley CA 94720, USA\\
$^{3}$Department of Physics and Astronomy, University of Manchester, Sackville 
      Street, P.O.Box 88, Manchester M60 1QD, UK}

\date{Submitted 2006}

\pagerange{\pageref{firstpage}--\pageref{lastpage}}

\pubyear{2006}

\begin{document}

\renewcommand{\topfraction}{1.}
\renewcommand{\bottomfraction}{1.}
\renewcommand{\textfraction}{0.}

\date{Accepted 2006 August 18. Received 2006 July 28; in original form 2006 June
5}
\maketitle
\label{firstpage}
\begin{abstract}
We investigate the conditions of star formation in the Large Magellanic Cloud. 
We have conducted a survey for water maser emission arising from massive young 
stellar objects in the 30 Doradus region (N\,157) and several other H\,{\sc ii}
regions in the Large Magellanic Cloud (N\,105A, N\,113 and N\,160A). We have
identified a new maser source in 30\,Dor at the systemic velocity of the LMC. 
We have obtained 3$-$4\,$\mu$m spectra, with the ESO Very Large Telescope, of 
two candidate young stellar objects. N\,105A\,IRS1 shows H recombination line 
emission and its Spectral Energy Distribution (SED) and mid-infrared colours are
consistent with a massive young star ionising the molecular cloud. N\,157B\,IRS1
is identified as an embedded young object, based on its SED and a tentative 
detection of water ice. The data on these four H\,{\sc ii} regions are combined 
with mid-infrared archival images from the Spitzer Space Telescope to study the 
location and nature of the embedded massive young stellar objects and signatures
of stellar feedback. Our analysis of 30\,Dor, N\,113 and N\,160A confirms the 
picture that the feedback from the massive O and B-type stars, which creates the
H\,{\sc ii} regions, also triggers further star formation on the interfaces of 
the ionised gas and the surrounding molecular cloud. Although in the dense cloud
N\,105A star formation seems to occur without evidence of massive star feedback,
the general conditions in the LMC seem favourable for sequential star formation 
as a result of feedback. In an appendix we present water maser observations of 
the galactic red giants R\,Doradus and W\,Hydrae.
\end{abstract}


\section{Introduction}

One of the great unknowns in our understanding of star formation concerns the 
role played by galactic environmental parameters. For instance, molecular cloud 
processes, e.g. cooling and magnetic field diffusion, depend on the presence of
metals, thus it is unlikely star formation is not affected by metallicity. The 
Magellanic Clouds have metallicity and density of the interstellar media (ISM) 
that are lower than in the Milky Way. Thus, they provide a star formation 
template that is more representative of the formation of stars at higher 
redshifts and offer a step closer towards exploring the formation of the first 
generation of stars at zero metallicity. 

Young pre-main-sequence populations have recently been identified in H\,{\sc ii}
regions in the Magellanic Clouds \citep{nota06,gouliermis06}. However, 
relatively little work has been done in investigating the earlier, more 
embedded stages of star formation in the Magellanic Clouds. At the distance of 
the Large Magellanic Cloud (LMC, $\sim$\,50\,kpc), one is limited to study only
the most massive stellar embryos that will form massive O and B stars. Infrared
(IR) observations are an excellent way to investigate the properties of such 
objects, both because they are heavily embedded and thus invisible at shorter 
wavelengths and also because they are surrounded by dense envelopes of gas, dust
and ices that are revealed at these wavelengths \citep{vandishoeck04}. Recently,
a few massive young stellar objects (YSOs) in the LMC have been investigated 
\citep{vanloon05b,jones05}.

Molecular masers (in particular water, hydroxyl and methanol) are very bright 
emission lines that seem to be closely associated with the earliest stages of 
massive star formation \citep{debuizer05} when the YSOs are difficult to 
detect even at IR wavelengths. Maser emission is found in the vicinity of 
embedded YSOs and thus they are powerful beacons of current star formation in a 
molecular cloud. Several water maser sources have been discovered in the 
Magellanic Clouds (see below). In this contribution we describe a new survey of 
water masing sources in the LMC at 22\,GHz with the Parkes Telescope.

In recent years, the 30 Doradus Nebula (N157) has become the paradigm of large 
scale triggered star formation. As the largest and nearest extragalactic 
H\,{\sc ii} region in the Local Group it offers a rare insight into the spatial 
and temporal properties of starbursts. \citet{walborn99} identified the 
so-called star-formation fronts, the interfaces between the energetic outflows 
from the central compact cluster R\,136 (first generation star formation 
episode) and the surrounding molecular clouds, in which there is evidence of 
ongoing triggered star formation. Water maser emission has also been detected at
various locations in 30\,Dor \citep{whiteoak83,vanloon01,lazendic02}. The wealth
of bright protostars identified in the region \citep{rubio98,brandner01} 
hints at the possibility that more masing activity hitherto undetected might be
present. We describe our mapping survey of the inner region of 30\,Dor, in 
which we have identified 4 distinct maser sources, one of them new. 

Three other H\,{\sc ii} regions in the LMC, known to show water maser emission 
\citep[N\,113, N\,105A and N\,160A;][]{whiteoak83,whiteoak86} are also observed 
at 22\,GHz, as well as most of the previous non-detections for water maser 
emission in the LMC and SMC \citep{scalise82,whiteoak83}. Since our
observations, \citet{lazendic02} also reported 22\,GHz water maser emission
towards N\,159.

The H\,{\sc ii} region N\,113 \citep{henize56} shows a complex structure of
H$\alpha$ bubbles with a rich molecular gas and dust morphology. Several young 
clusters are associated with N\,113 \citep*{bica92} and current star formation 
activity is occuring within continuum sources in the central area of the nebula 
\citep{brooks97,wong06}. The H\,{\sc ii} region N\,105 \citep{henize56} is a 
complex of evacuated bubbles and dense molecular material with several young 
clusters associated with it \citep{ambrocio98}; current star formation as 
indicated by maser activity seems to concentrate in the denser central part of 
the region, N\,105A. N\,160A is the brightest component in the N\,160 
H\,{\sc ii} complex \citep{henize56}. Besides maser emission, both an embedded 
protostar \citep{henning98} and a dense molecular core \citep{bolatto00} point 
at ongoing star formation. \citet{malayeri02} and \citet{nakajima05} describe 
the gas morphology and stellar content of N\,160.

In order to investigate the relationship between the masers and the gas 
kinematics and star formation activity, we have obtained narrow-band H$\alpha$ 
images at the AAT of N\,113, and L$^\prime$-band images and 3$-$4\,$\mu$m 
spectroscopy at the ESO-VLT in N\,157B and N\,105A. These new observations, 
combined with archival IRAC/Spitzer images are used to relate the embedded 
population to the local gas and dust morphology. 

\section{Observations}

The new observations used in this paper are described in this section and a
summary is given in Table\,\ref{overview}.
\begin{table*}
\caption[]{Overview of the new data on the H\,{\sc ii} regions discussed in
detail in this paper.}
\label{overview}
\begin{tabular}{lcccccc}
\hline\hline
H\,{\sc ii} region&22 GHz &H$\alpha$ image&$L^\prime$ image &Spitzer 8\,$\mu$m image &L-band spectrum&IR SED\\
\hline
30\,Dor (N\,157A) &map            & & &+&           &    \\
N\,157B           &single pointing& &+&+&IRS1       &IRS1\\
N\,113            &sparse map     &+& &+&           &    \\
N\,105A           &single pointing& &+&+&IRS1, Blob &IRS1\\
N\,160A           &single pointing& & &+&           &    \\
\hline
\end{tabular}
\end{table*}

\subsection{The 22 GHz survey at Parkes}

\label{parkes}

The 64\,m radio telescope (effective diameter of 45\,m) at Parkes, Australia,
was used from June 30 to July 9 2001, with the K-band (1.4 cm) receiver plus
autocorrelator back-end to observe the $6_{16} \rightarrow 5_{23}$ rotational
transition of ortho-H$_2$O at a rest frequency of 22.23507985 GHz. The
observations were performed at $\sim22.216$\,GHz yielding a velocity coverage
of $\sim860$\,km\,s$^{-1}$ with 0.42\,km\,s$^{-1}$\,channel$^{-1}$. Using the 
dual circular feed, spectra were obtained simultaneously in left and right 
circular polarisation; these were then averaged.

The system temperature varied between 120 and 140\,K. The conversion factor from
antenna temperature to flux density was 6\,Jy\,K$^{-1}$. Observing conditions 
were sometimes rather unstable due to clouds. The on-source integration time was
20\,min per pointing. The pointing and focus were checked regularly by observing
the bright maser sources R\,Doradus and W\,Hydrae (see Appendix\,\ref{rdor}). 
The absolute flux calibration is accurate to $\sim$\,20 per cent.

\begin{figure*}
\includegraphics[width=180mm]{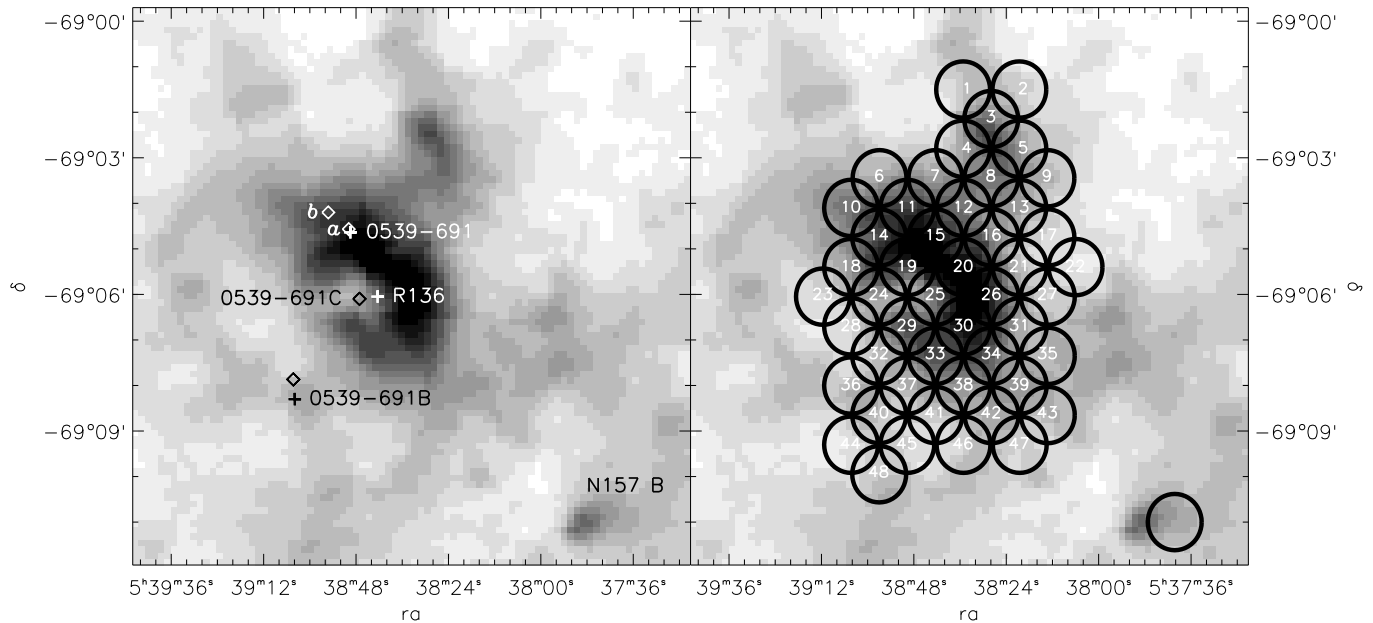}
\caption[]{22 GHz pointings in 30\,Dor superimposed on an MSX image at 8 $\mu$m.
On the left, the $+$ symbols identify R\,136 and the published positions of the
maser sources 0539$-$691 and 0539$-$691\,B; $\diamond$ symbols show the 
positions of the maser sources 0539$-$691 ({\it a} and {\it b}), 0539$-$691B and
0539$-$691C determined in this paper (Section\,\ref{maser_30dor}). On the
right, the positions observed are schematically represented and numbered. An
extra pointing was performed towards N\,157B, in the lower right corner of the
image. The greyscale is logarithmic between 10$^{-10}$ and $3 \times 10 ^{-9}$
W\,cm$^{-2}$\,sr$^{-1}$.}
\label{pointings}
\end{figure*}

The obtained spectra were corrected for two distinct baseline effects: a 
low-frequency feature (easily removed with a 3$^{\rm rd}$-degree polynomial) and
an interference signal with a frequency of $\sim2.85$ MHz. The shape of this 
interference was re-constructed by median averaging the cycles within each 
spectrum, and was then subtracted from the spectrum.

The observations in the central area of 30\,Doradus (LHA\,120-N\,157A) were 
performed on a double grid system ($x,y$): the primary grid has a separation of 
one beam, i.e.\ $1.3\arcmin$; the secondary grid positions are obtained by 
shifting the primary grid positions by $39\arcsec$ in right ascension and 
declination. We performed 58 pointings at 47 distinct positions around R\,136 
(some positions deemed interesting were observed twice), covering a 
$5\arcmin\times5\arcmin$ area. The pointings are shown in Fig.\,\ref{pointings},
superimposed on a Midcourse Space eXperiment (MSX) 8\,$\mu$m image. An extra 
pointing was performed on N\,157B, but note that it is offset by 1$\arcmin$ 
from the peak emission in the mid-IR.

The observations of N\,113 were performed in a 5-position dither pattern with 
$39\arcsec$ shifts. Single pointings were performed towards a number of other 
H\,{\sc ii} regions; maser sources were detected towards N\,105A and N\,160A, 
while for 11 other regions no masing source was detected (see 
Appendix\,\ref{other_masers}).

\subsection{The 3$-$4\,$\mu$m observations at the VLT}

The Infrared Spectrometer And Array Camera (ISAAC) on the European Southern 
Observatory (ESO) Very Large Telescope (VLT), Chile, was used on 7 and 8 
December 2003 to obtain long-slit spectra between 2.85 and 4.15 $\mu$m of two 
candidate embedded YSOs. MSX\,LMC\,888 \citep*{egan01} 
(henceforth N\,157B\,IRS1) is located in the 30\,Dor region and was selected 
based on its very red colours (consistent with a massive YSO) while 
MSX\,LMC\,80 (N\,105A\,IRS1) is a candidate protostar first identified by 
\citet*{epchtein84} --- see Sect.\,\ref{vlt} for a complete discussion on these 
sources.

The resolving power of $\lambda/\Delta\lambda\sim700$ was set by the 
$\sim0.5^{\prime\prime}$ seeing rather than the $2^{\prime\prime}$ slit width.
The thermal-IR technique of chopping and nodding was used to remove the
background, with a throw of $10^{\prime\prime}$, jittering within
$2^{\prime\prime}$ to correct for bad pixels. Total exposure times were 12
minutes. The spectra were extracted using an optimal extraction algorithm. An
internal Xe+Ar lamp was used for wavelength calibration. The relative spectral
response was calibrated by dividing by the spectrum of the B-type star
HIP\,020020, observed within an hour from the targets at the same airmass and
similar seeing. This removed most of the telluric absorption lines --- except 
the saturated methane line at 3.32 $\mu$m ---  but introduced artificial 
emission features due to the photospheric lines of HIP\,020020. The spectra were
therefore multiplied by a hot blackbody continuum with Gaussian-shaped
absorption lines of Br$\alpha$ 4.052, Pf$\gamma$ 3.741, Pf$\delta$ 3.297,
Pf$\epsilon$ 3.039 and Pf$\zeta$ 2.873 $\mu$m scaled to match those in
HIP\,020020. More details can be found in \citet{vanloon06}.

The acquisition images were obtained through the L$^\prime$ filter 
(3.78\,$\mu$m, bandwidth ${\Delta}\lambda=0.58$\,$\mu$m). At a scale of
$0.148^{\prime\prime}$\,pixel$^{-1}$, the (array-windowed) field-of-view was
$57^{\prime\prime}\times57^{\prime\prime}$. The images were obtained in
chopping-only mode, with a throw of $10^{\prime\prime}$ in the North-South
direction. The total exposure time was 20 seconds. The photometric accuracy is
of the order of 0.1$-$0.2\,mag, found from cross-correlation of acquisition
images of other targets in the same observing run with dedicated
L$^\prime$-band photometry \citep{vanloon05a}.

\subsection{The H$\alpha$\ mapping at the Anglo-Australian Telescope}

The Taurus Tunable Filter instrument on the Anglo-Australian Telescope (AAT),
Australia, was used on 17 July 2001 to obtain narrow-band H$\alpha$ images of
N\,113. The blue etalon was used in combination with the EEV CCD, at a
spectral resolution of $\Delta\lambda\sim$7\,\AA\ around H$\alpha$. The central
wavelength was set to $\lambda_0=6568$\,\AA, corresponding to a Doppler shift
similar to that of the LMC. The technique of charge-shuffling was applied to
take a series of three consecutive images before reading out the CCD, where
each image was taken on the same area of the CCD but at a central wavelength
of $\lambda_0+\Delta\lambda$, $\lambda_0$ and $\lambda_0-\Delta\lambda$,
respectively. Each image received an integration time of 1 minute, and a total
of six such cycles were performed before CCD read-out. This was repeated with
images taken at $\lambda_0+3\Delta\lambda/2$, $\lambda_0+\Delta\lambda/2$ and
$\lambda_0-\Delta\lambda/2$, respectively, to improve the spatial and spectral 
sampling of H$\alpha$; as the instrument is placed in the pupil the wavelength
changes with radial distance $r$ in pixels as \citep[cf.][]{bland89}:
\begin{equation}
\frac{\lambda}{\lambda_0}=1-\frac{r^2}{2}\left(\frac{p}{f}\right)^2,
\end{equation}
where $p=13.5$ $\mu$m is the pixel size and $f=127.8$ mm the camera focal 
length. The two sets of observations were repeated with an offset of $1^\prime$ 
in the North-South direction to cover more of N\,113. The observations were made
at a distance of $50^\circ$ from the zenith and a seeing of 
$1.8^{\prime\prime}$.

The data were corrected for bias offset and divided by a flatfield image. For
each of the two pointings the images were combined to produce one image
corresponding to the H$\alpha$ line emission and another image corresponding
to the underlying continuum emission. This was done on a pixel-by-pixel basis
by assuming that the H$\alpha$ emission has a gaussian spectral shape with
a FWHM of 7 \AA, and that the continuum is flat. The resulting H$\alpha$ image
is very clean with most stars removed. The main H$\alpha$ emission structures 
are so strong that they are still present in the ``continuum'' image, albeit at 
a much lower level (Sect.\,\ref{halpha}). 

\subsection{Additional infrared data}
\label{add_ir}

Near-IR $JHK_{\rm s}$ photometry is taken from the 2-Micron All-Sky Survey
\citep[2MASS,][]{cutri03}. Embedded massive stars have very red $J-K_{\rm s}$ 
colours, thus, at the distance of the Magellanic Clouds, most embedded YSOs
are only detected by 2MASS in the K$_{\rm s}$-band. Hence, bright 
K$_{\rm s}$-band sources with faint or undetected shorter wavelength 
counterparts in 2MASS are embedded YSO candidates.

For the two luminous IR objects, N\,157B\,IRS1 and N\,105A\,IRS1 
(Sect.\,\ref{vlt}), mid-IR photometry at 8.28, 12.1, 14.7 and 21.3\,$\mu$m is 
taken from version 2.3 of the Mid-course Space eXperiment (MSX) Point Source 
Catalogue \citep{egan03}. The spatial resolution of these data varies from 
$7^{\prime\prime}$ at 8.28\,$\mu$m, to $18^{\prime\prime}$ at 21.3\,$\mu$m. We 
also collected scans from the IRAS data server\footnotemark 
\footnotetext{http://www.astro.rug.nl/IRAS-Server/} for these two sources, to 
measure their 12, 25, 60 and 100 $\mu$m flux densities using the Groningen 
Image Processing SYstem (GIPSY) software with the {\sc scanaid} tool. These 
measurements were fully consistent with the IRAS Point Source Catalogue 
\citep{beichman88}, but contrary to the PSC we were able to obtain a 
measurement for N\,157B\,IRS1 also at 60 and 100\,$\mu$m. In addition, for 
N\,105A\,IRS1 we inspected Low Resolution Spectrograph scans from the IRAS 
data server. The three complete scans are fully consistent with each other, 
and show a weak depression around 10\,$\mu$m which could be indicative of 
absorption by silicate dust. The mid-IR photometry helps constrain the 
luminosity and hence the mass of the star.

In our analysis of the morphology of the 4 H\,{\sc ii} regions we also made use
of archival images obtained with the Spitzer Space Telescope \citep{werner04}, 
using its Infrared Array Camera \citep[IRAC,][]{fazio04}. IRAC images of 
30\,Dor were obtained under Early Release Observation programme \#1032 (P.I. 
B. Brandl). Both N\,105A and N\,113 observations are part of the cycle 2 
Legacy programme \#20203 (P.I. M. Meixner) while the N\,160A observations are 
part of the Guaranteed Time Observations programme \#124 (P.I. R. Gehrz). 
Extended pipeline products (i.e. flux-calibrated image mosaics) for the 4 IRAC 
bands (3.6, 4.5, 5.8 and 8.0\,$\mu$m) were retrieved from the Spitzer 
archive\footnotemark \footnotetext{http://ssc.spitzer.caltech.edu/archanaly/}. 
We have performed aperture photometry on the IRAC mosaics, for the sources 
present in the L$^\prime$ acquisition images. We performed aperture corrections 
but not array-location dependent corrections or colour corrections 
\citep{reach05}, so fluxes might be uncertain by as much as 10 per cent, still 
fully adequate for our analysis (see Sect.\,\ref{seds}).

\section{Results}

\subsection{22 GHz survey}

\subsubsection{Mapping of 30\,Dor (N\,157)}
\label{maser_30dor}

\begin{figure}
\includegraphics[width=84mm]{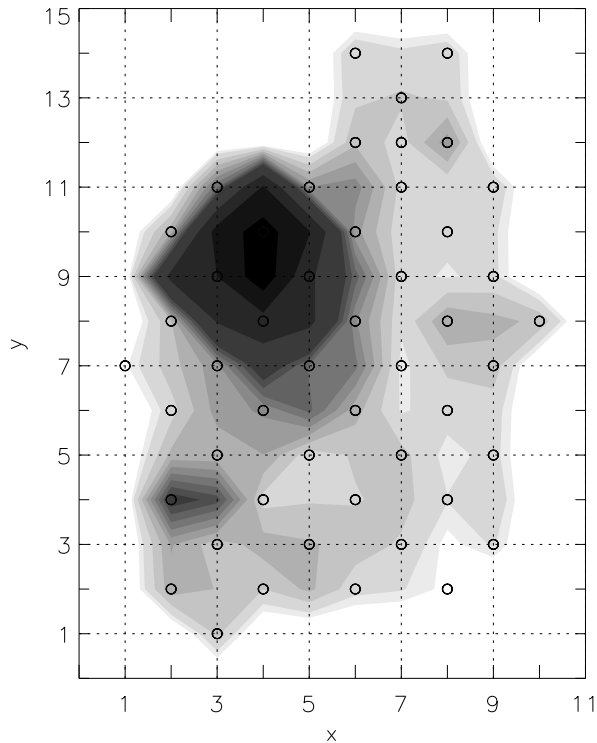}
\caption[]{Intensity map of the 22 GHz survey of 30\,Dor. In this contour
plot, the darker regions indicate possible masing sources. Open circles
indicate the grid positions that were observed, corresponding to the pointings
in Fig.\,\ref{pointings}. Position (1,1) corresponds to (RA,Dec)\,=\,($05^{\rm 
h}39^{\rm m}11.6^{\rm s}$, $-69\degr09\arcmin57\arcsec$), and the grid 
spacing is half a beam (39$\arcsec$). The intensity scale in this plot is 
from $\lsim 1$ to 5\,Jy\,km\,s$^{-1}$. Comparison with Fig.\,\ref{pointings} 
suggests that maser emission seems to concentrate towards the brighter areas of 
the nebula at IR wavelengths.}
\label{intensity_map}
\end{figure}

The goal of the 30\,Dor survey was to detect maser sources and locate their
positions to better than the beam size. We assume that the sensitivity of the 
telescope beam is a function of the distance to beam centre, represented by a 
gaussian distribution. We reconstruct the spectra at each position in the 
observed grid by combining each spectrum with the available adjacent spectra in
the grid, weighted by their variance and the gaussian weight corresponding to 
the distance between the grid points. With this process we improve 
signal-to-noise ratio and detection sensitivity. By integrating the flux of the 
strongest spectral feature in each spectrum, we construct the intensity map 
shown in Fig.\,\ref{intensity_map}. This map shows the location of the strongest
intensity peaks; in a second step the spectra at each of those intensity peaks 
and closely surrounding ones are checked (e.g., for velocity consistency) to 
separate bona-fide maser sources from spurious peaks. The detection sensitivity
is $\sim$\,1\,Jy\,km\,s$^{-1}$ ($\sim$\,5$\sigma$). It can been seen from this 
image that there are 2 main flux enhancement regions: a complex region towards 
the top, that actually includes multiple masing sources and a single source area
towards the bottom of the diagram. 

\begin{table*}
\caption[]{Observed properties of water maser detections: peak velocity, peak 
intensity, source position and pointings identification (for 30\,Dor only, 
Fig.\,\ref{pointings}). We estimate that the positions in 30\,Dor are accurate 
to about a quarter of a beam size, approximately 20$\arcsec$. 0539$-$691C is a 
hitherto unknown component. For the other sources we are not able to compute 
accurate positions; for these, telescope pointing positions are listed. No maser 
emission was detected towards N\,157B and thus we provide the rms level in the 
spectrum.}
\label{maser_positions}
\begin{tabular}{lccccl}
\hline
Maser source&peak velocity &peak intensity  &RA 2000   &Dec 2000   &pointings\\
            &(km\,s$^{-1}$)&(Jy)&($^{h\,\,m\,\,s}$)&($\degr\,\,\arcmin\,\,\arcsec$)&identification \\
\hline
0539$-$691{\it a}  &269.5  &2.75$\pm$0.60            &05 38 49.9&$-$69 04 34\rlap{$\dagger$}&6,\,7,\,10,\,11,\,12,\,14,\,15,\,18,\,19\\
0539$-$691{\it b}  &261.0  &0.38$\pm$0.09            &05 38 55.2&$-$69 04 12&6,\,10,\,11,\,18,\,19 \\
0539$-$691B        &194.0  &0.36$\pm$0.09            &05 39 04.3&$-$69 07 52&32,\,36 \\
0539$-$691C        &266.0  &1.01$\pm$0.27            &05 38 47.1&$-$69 06 06&19,\,20,\,25,\,29,\,30\\
\hline
N\,113a            &253.0  &\llap{7}3.8$\pm$16.\rlap{0}    & 05 13 23.1 & $-$69 22 34& \\
N\,113b            &254.5  &\llap{1}9.4$\pm$4.4 & --- & ---& \\
N\,113c            &251.0  &  4.6$\pm$1.1 & --- & ---& \\
N\,113d            &258.0  &  2.0$\pm$0.5 & --- & ---& \\
\hline
N\,160A            &253.0  &  3.3$\pm$0.7 & 05 39 42.7 & $-$69 38 26\\
N\,105A            &260.0  &  1.5$\pm$0.3 & 05 09 50.7 & $-$68 53 23\\
\hline
N\,157B            &---    &0.13 (rms)    & 05 37 40.2 & $-$69 11 00\rlap{$\ddagger$}\\
\hline
\end{tabular}

\flushleft $\dagger$ a more accurate position for this source can be found in \citet{lazendic02}.
\vspace{-2mm} \flushleft $\ddagger$ this position is 1$\arcmin$ to the West of the peak IR emission.
\end{table*}

\begin{figure}
\includegraphics[width=84mm]{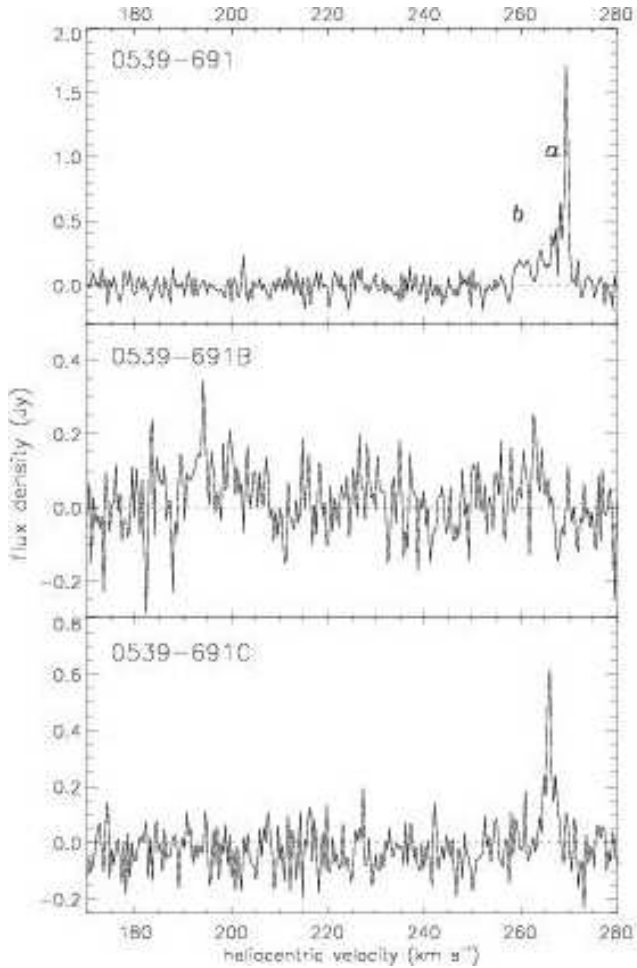}
\caption[]{Maser sources detected in 30\,Doradus. The spectra correspond to
the positions 0539$-$691, 0539$-$691B and 0539$-$691C respectively. The masing
structure at 0539$-$691 (pointing 11) has two distinct components ({\it a} and 
{\it b} as indicated) at velocities consistent with the local ISM velocity. The 
maser at 0539$-$691B (pointing 36) has a strongly supersonic velocity as 
observed by \citet{vanloon01}. The new maser source 0539-691C (pointing 25) 
has a single component at the systemic velocity.}
\label{maser_components}
\end{figure}

On kinematic and flux intensity grounds, we have identified 4 distinct water 
maser peaks in 30\,Doradus. In each spectrum where the maser emission is 
present, we measure the peak intensity of each component at the velocities 
listed in Table\,\ref{maser_positions}. We obtain a spatial representation of 
the intensity of each component that we fit with a two-dimensional gaussian 
distribution with three free parameters: the amplitude and position ($x,y$) of 
the centre --- the width of the distribution is fixed by the beam size. The 
amplitude translates into the intrinsic intensity of the source and the centre 
is converted into right ascension and declination coordinates of each component,
with a typical positional accuracy of a quarter of a beam size, 
$\sim$\,20$\arcsec$. In Table\,\ref{maser_positions} we list the measured 
velocity, intensity and position of each maser source, as well as their 
identifications (see below). The last column provides the pointings at which the
masers were detected (Fig.\,\ref{pointings}). The maser spectra are presented in
Fig\,\ref{maser_components}.

0539$-$691 was first detected at 22\,Ghz by \citet{whiteoak83}, in a region 
$\sim1.5\arcmin$ northeast of R\,136. \citet{vanloon01} further analysed this 
region, detecting 0539$-$691 as well as a new water maser source, 0539$-$691B,
located $\sim2.5\arcmin$ southeast of R\,136. On kinematic grounds the authors 
distinguish two masing sources in 0539$-$691, one with a velocity consistent 
with the systemic velocity of the local ISM ($v_{\rm hel}\sim270$\,km\,s$^{-1}$)
--- this is the source originally identified by \citet{whiteoak83} and also 
observed by \citet{lazendic02} --- and a component blue-shifted by 
$\sim$\,70\,km\,s$^{-1}$. 0539$-$691B was also found to be blueshifted with 
respect to the systemic velocity by $\sim$\,90\,km\,s$^{-1}$. 

We find that the maser 0539$-$691 comprises 2 components ({\it a} and {\it b}) 
with peak velocities $\sim261$ and 270\,km\,s$^{-1}$. They are very close 
($\sim1.0\arcmin$) and component {\it b} is rather weak, thus it is 
conceivable that it is a single source. \citet{lazendic02} also detect a
component with peak velocity 269.5\,km\,s$^{-1}$ and they pinpoint its location
very accurately. They claim that their spectrum also contains additional faint
emission extending down to 258\,km\,s$^{-1}$ that might be related to our
component {\it b}.  We did not detect the blueshifted component detected in 
0539$-$691 by \citet{vanloon01}. We do identify a weak source with peak velocity
of $\sim$\,194\,km\,s$^{-1}$, located $\sim4.6\arcmin$ from 0539$-$691, with 
0539$-$691B \citep{vanloon01}. We have discovered a new maser source closer to 
R\,136 ($\sim2.8\arcmin$ South from 0539$-$691), that we call 0539$-$691C. It 
has a velocity of $\sim266$ km\,s$^{-1}$, consistent with the systemic velocity.
Despite the large survey area, this is the only new component we identified. No 
maser emission was detected towards N\,157B, at an rms level of 0.13\,Jy but 
note that the target was not well centered in the telescope beam.

\subsubsection{Sparse map of N\,113}

\begin{figure}
\includegraphics[width=84mm]{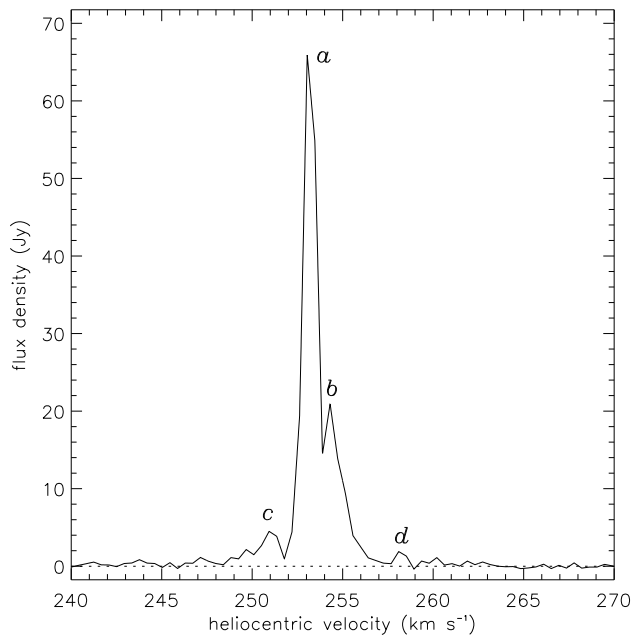}
\caption[]{Water maser sources detected in N\,113. This spectrum is the central
one from our 5-position sparce map (see Section\,\ref{parkes}). The different
components are labelled {\it a}, {\it b}, {\it c} and {\it d}.}
\label{maser_components_n113}
\end{figure}

\begin{figure}
\includegraphics[width=84mm]{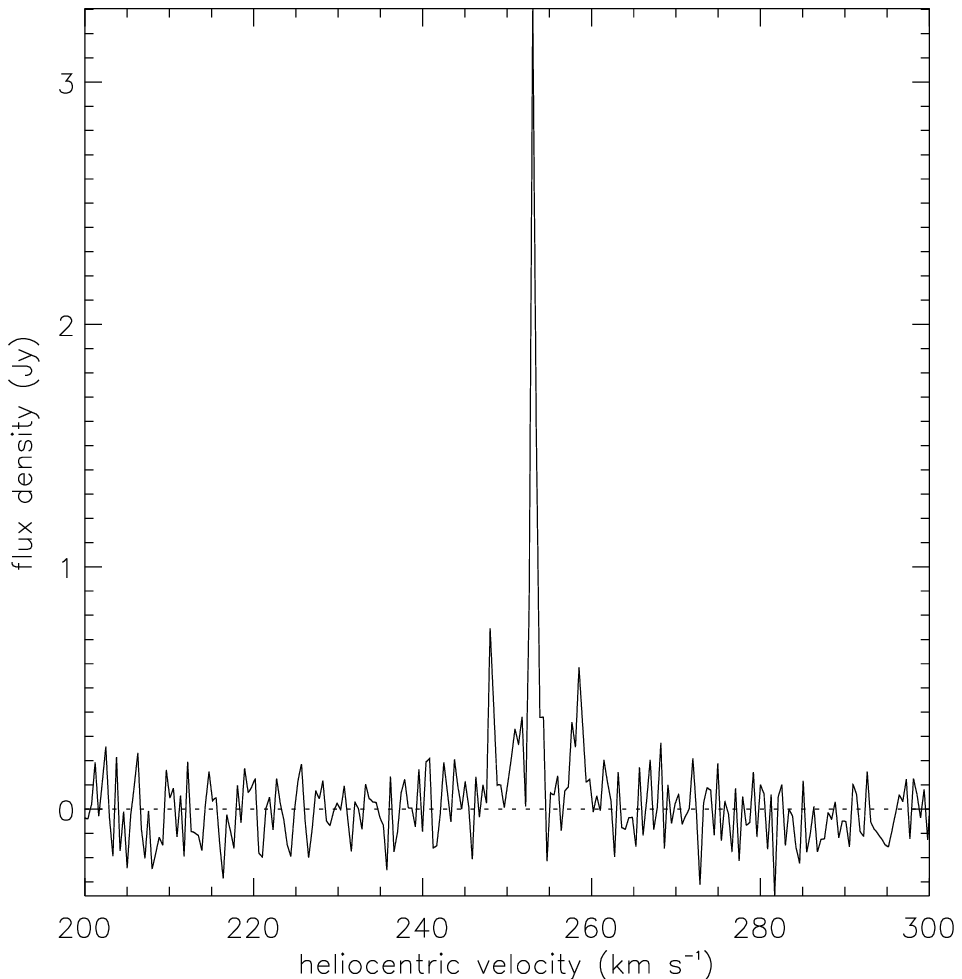}
\includegraphics[width=84mm]{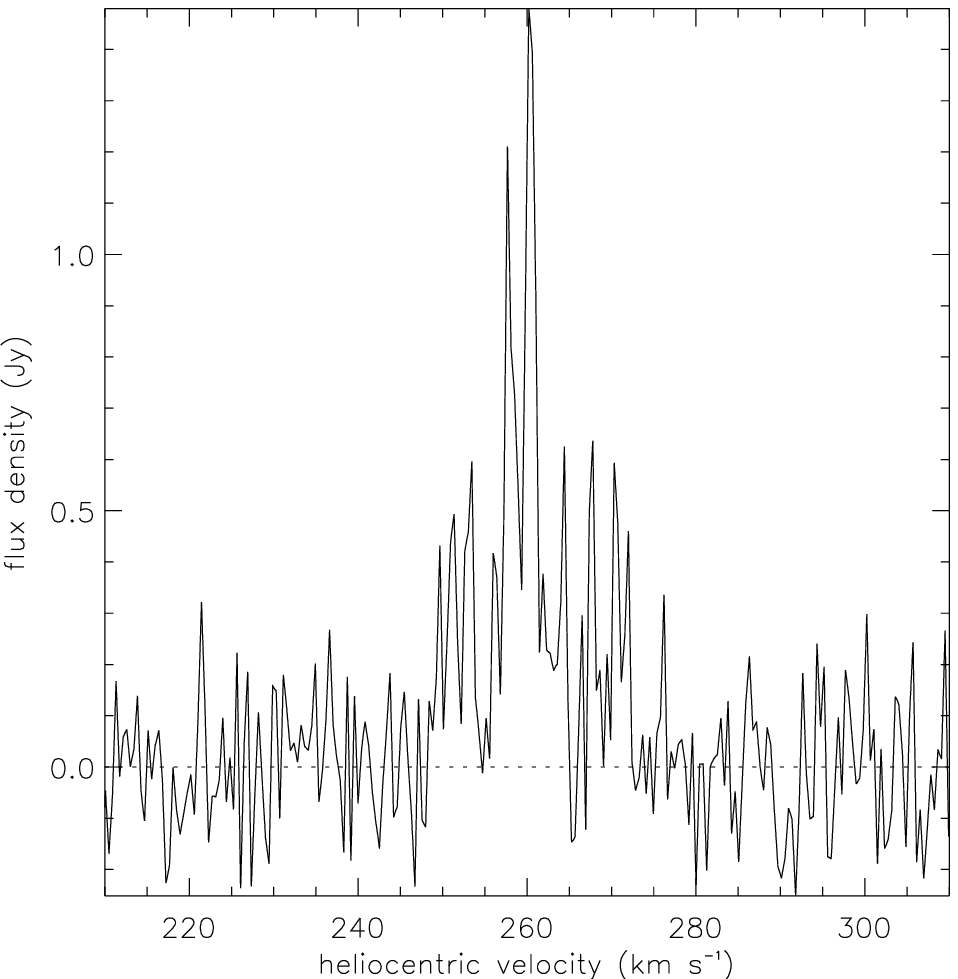}
\caption[]{Water maser sources detected towards N\,160A (top) and N\,105\,A 
(bottom).}
\label{maser_other}
\end{figure}

Water maser emission at 22\,GHz was first observed in N\,113 by
\citet{whiteoak86}. We performed 5 pointings in this region. The method used to 
analyse the spectra of N\,113 is similar to what was outlined in the previous 
section for 30\,Doradus, except with fewer pointings. In 
Fig.\,\ref{maser_components_n113}, we plot the central spectrum of our mosaic of
pointings in N\,113, with 4 components identified as {\it a}, {\it b}, {\it c} 
and {\it d}. 

We were not able to disentangle the positions of the different components, so 
we cannot establish positively how many spatially-distinct masers are there in 
this region. However, from the analysis of the peak intensity variations in our
mosaic of pointings, we find evidence that component {\it c} might originate 
from a different position than the remaining components: at the position 
northwest of the central pointing, the peak intensity of component {\it c} was 
reduced by $\sim60$ per cent, while the peak intensities of the other components
were reduced to $\sim96$ per cent. \citet{lazendic02} recently identified 2 
distinct masers sources in N\,113, $26\arcsec$ apart, and with peak velocities 
253 and 250\,km\,s$^{-1}$. We identify our multiple component source {\it a, b}
and {\it d} with the \citet{lazendic02} component at 253\,km\,s$^{-1}$; based on
the peak velocity, we believe our component {\it c} is the \citet{lazendic02} 
component at 250\,km\,s$^{-1}$. The strongest masing component is weaker than 
in the \citet{lazendic02} measurements. We note that water maser emission 
associated with YSOs is known to be variable both in intensity and velocity 
\citep{panagi93,tofani95}. 

\subsubsection{Other regions in the LMC}

Fig.\,\ref{maser_other} shows the maser detections towards N\,160A (top) and 
N\,105A (bottom). We are unable to derive accurate position information from these 
observations but both these sources have been observed also by 
\citet{lazendic02}, providing positions with subarcsecond accuracy. The maser 
spectrum of N\,105A is similar to that reported previously, both in kinematic 
complexity and intensity. For N\,160A two maser locations, approximately 
44$\arcsec$ apart have been identified previously at essentially the same 
velocity. It is very likely that both those components contribute to the signal
in our spectrum; indeed the flux ratios of the brighter component at 
253\,km\,s$^{-1}$ to the other two components at 248 and 259\,km\,s$^{-1}$ are 
consistent with what would be expected if the two components of 
\citet{lazendic02} were combined.

\subsection{VLT 3$-$4 $\mu$m observations}
\label{vlt}

\begin{figure}
\includegraphics[width=84mm]{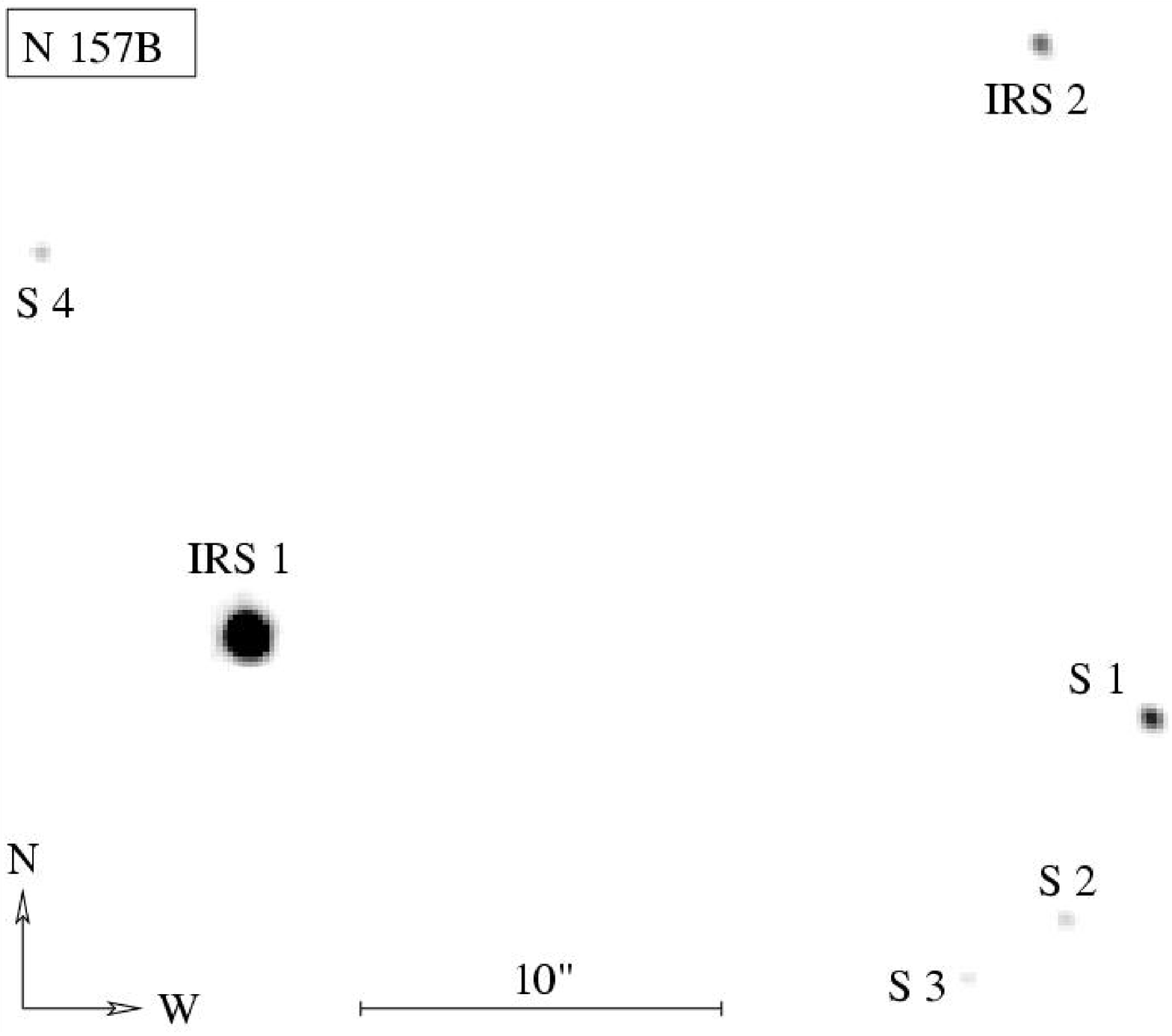}
\includegraphics[width=84mm]{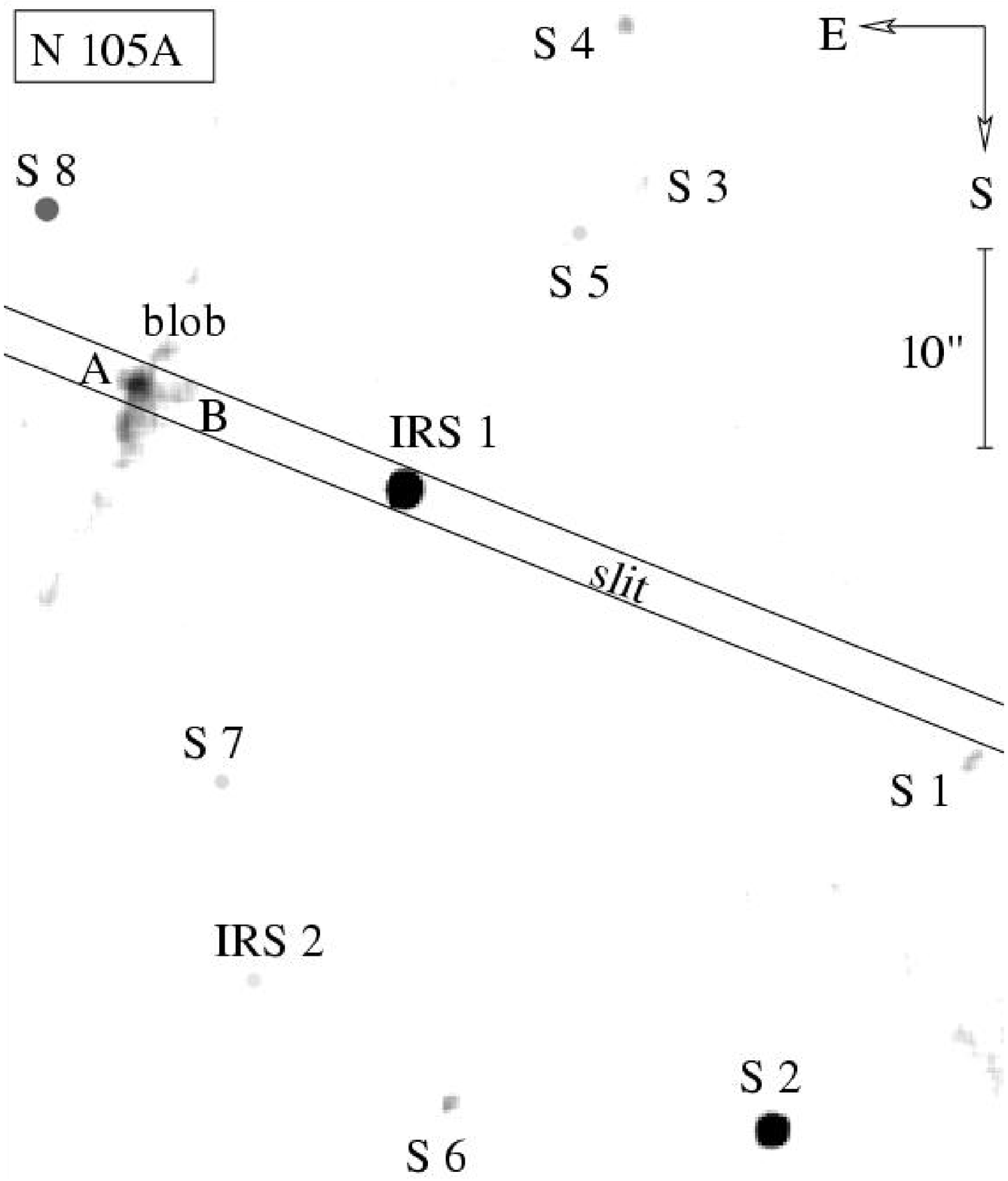}
\caption[]{L$^\prime$-band acquisition images of N\,157B (top) and N\,105A
(bottom), with identified stellar sources (``S'') and infrared sources 
(``IRS''). For N\,157B, the slit was centred on IRS1 and aligned N$-$S, while 
for N\,105A the slit orientation was chosen to include the diffuse source 
blob\,A (see text). The faintest stars in the images are $L^{'} = 12.76$\,mag 
(N157B\,S3) and $L^{'} = 14.30$\,mag (N105A\,IRS2), respectively.}
\label{vlt_l}
\end{figure}

\begin{table*}
\caption[]{List of objects in N\,157B and N\,105A for which we have 
3$-$4\,$\mu$m data, in order of increasing RA (all J2000 coordinates are based 
on 2MASS). Near-IR magnitudes are from 2MASS (JHK$_{\rm s}$) and our 
L$^\prime$-band acquisition images.}
\label{nir_t}
\begin{tabular}{lcccccccc}
\hline
LHA\,120-...  &RA 2000           &Dec 2000          & J   & H   & K$_{\rm s}$ & L$^\prime$&J$-$K$_{\rm s}$ & K$_{\rm s}-$L$'$\\ 
              &($^{h\,\,m\,\,s}$)&(${\degr,\,\arcmin\,\,arcsec}$)&(mag)&(mag)&(mag)        &(mag)&(mag)&(mag)\\
\hline
N\,157B\,S1   &05 37 45.5 & $-$69 11 10 & $13.14\pm0.05$ & $12.07\pm0.04$ & $11.58\pm0.03$ & $11.11\pm0.01$&1.56$\pm$0.06&0.47$\pm$0.03\\
N\,157B\,S2   &05 37 46.0 & $-$69 11 15 & $>13.2$	 & $>12.6$	  & $12.95\pm0.05$ & $12.45\pm0.06$&$>0.3$&0.50$\pm$0.08\\
N\,157B\,IRS2 &05 37 46.1 & $-$69 10 50 & $>17.0$	 & $>15.6$	  & $14.33\pm0.09$ & $11.39\pm0.02$&$>2.7$&2.94$\pm$0.09\\
N\,157B\,S3   &05 37 46.4 & $-$69 11 17 & $13.35\pm0.04$ & $13.03\pm0.06$ & $12.89\pm0.06$ & $12.76\pm0.07$&0.46$\pm$0.07&0.13$\pm$0.09\\
N\,157B\,IRS1 &05 37 50.3 & $-$69 11 07 & $15.93\pm0.12$ & $14.22\pm0.07$ & $11.45\pm0.03$ & \hspace{1.5mm}$7.57\pm0.01$&4.48$\pm$0.12&3.88$\pm$0.03\\
N\,157B\,S4   &05 37 51.4 & $-$69 10 56 & $13.81\pm0.05$ & $12.79\pm0.03$ & $12.50\pm0.04$ & $12.27\pm0.05$&1.31$\pm$0.06&0.23$\pm$0.06\\
\hline
LHA\,120-...    &RA 2000           &Dec 2000          & J   & H   & K$_{\rm s}$ & L$^\prime$&J$-$K$_{\rm s}$ & K$_{\rm s}-$L$'$ \\ 
                &($^{h\,\,m\,\,s}$)&($^{d\,\,m\,\,s}$)&(mag)&(mag)&(mag)        &(mag)&(mag)&(mag)\\
\hline
N\,105A\,S1     &05 09 45.6&$-$68 53 19&$14.31\pm0.03$ &$13.41\pm0.03$ &$13.28\pm0.04$ &$13.31\pm0.16$&1.03$\pm$0.05&\llap{$-$}0.03$\pm$0.16 \\
N\,105A\,S2     &05 09 47.5&$-$68 53 36&$11.68\pm0.02$ &$10.87\pm0.02$ &$10.63\pm0.02$ &$10.41\pm0.01$&1.05$\pm$0.03&0.22$\pm$0.02\\
N\,105A\,S3     &05 09 48.4&$-$68 52 51&$14.65\pm0.08$ &$13.84\pm0.04$ &$13.70\pm0.09$ &$13.43\pm0.14$&0.95$\pm$0.12&0.27$\pm$0.17\\
N\,105A\,S4     &05 09 48.5&$-$68 52 44&$13.89\pm0.05$ &$13.08\pm0.04$ &$12.98\pm0.05$ &$12.78\pm0.07$&0.91$\pm$0.07&0.20$\pm$0.09\\
N\,105A\,S5     &05 09 49.1&$-$68 52 54&$14.67\pm0.05$ &$13.90\pm0.07$ &$13.89\pm0.07$ &$13.74\pm0.17$&0.78$\pm$0.09&0.15$\pm$0.18\\
N\,105A\,S6     &05 09 50.3&$-$68 53 35&$15.42\pm0.09$ &$14.06\pm0.06$ &$13.73\pm0.07$ &$13.17\pm0.12$&1.69$\pm$0.11&0.56$\pm$0.14\\
N\,105A\,IRS1   &05 09 50.6&$-$68 53 05&$>15.3$        &$>14.7$        &$13.77\pm0.10$ &$\hspace{1.5mm}9.88\pm0.01$&$>1.5$&3.89$\pm$0.10 \\
N\,105A\,IRS2   &05 09 52.0&$-$68 53 29&&&&$14.30\pm0.34$ &&\\
N\,105A\,S7     &05 09 52.3&$-$68 53 20&$>15.9$        &$15.25\pm0.14$ &$14.16\pm0.15$ &$13.61\pm0.19$&$>1.7$&0.55$\pm$0.24 \\
N\,105A\,blob\,A&05 09 52.9&$-$68 53 01&$14.53\pm0.09$ &$13.88\pm0.11$ &$13.02\pm0.10$ &$12.83\pm0.11$&1.51$\pm$0.13&0.19$\pm$0.15\\
N\,105A\,blob\,B&05 09 52.5&$-$68 53 02&&&&&&\\
N\,105A\,S8     &05 09 53.8&$-$68 52 53&$13.40\pm0.04$ &$13.21\pm0.04$ &$12.88\pm0.05$ &$12.35\pm0.10$&0.52$\pm$0.06&0.53$\pm$0.11\\
N\,105A\,S9     &05 09 55.6&$-$68 53 13&$14.39\pm0.04$ &$13.34\pm0.04$ &$13.11\pm0.03$ &$12.96\pm0.17$&1.28$\pm$0.05&0.15$\pm$0.17\\
\hline
\end{tabular}
\end{table*}

\begin{figure}
\includegraphics[width=84mm]{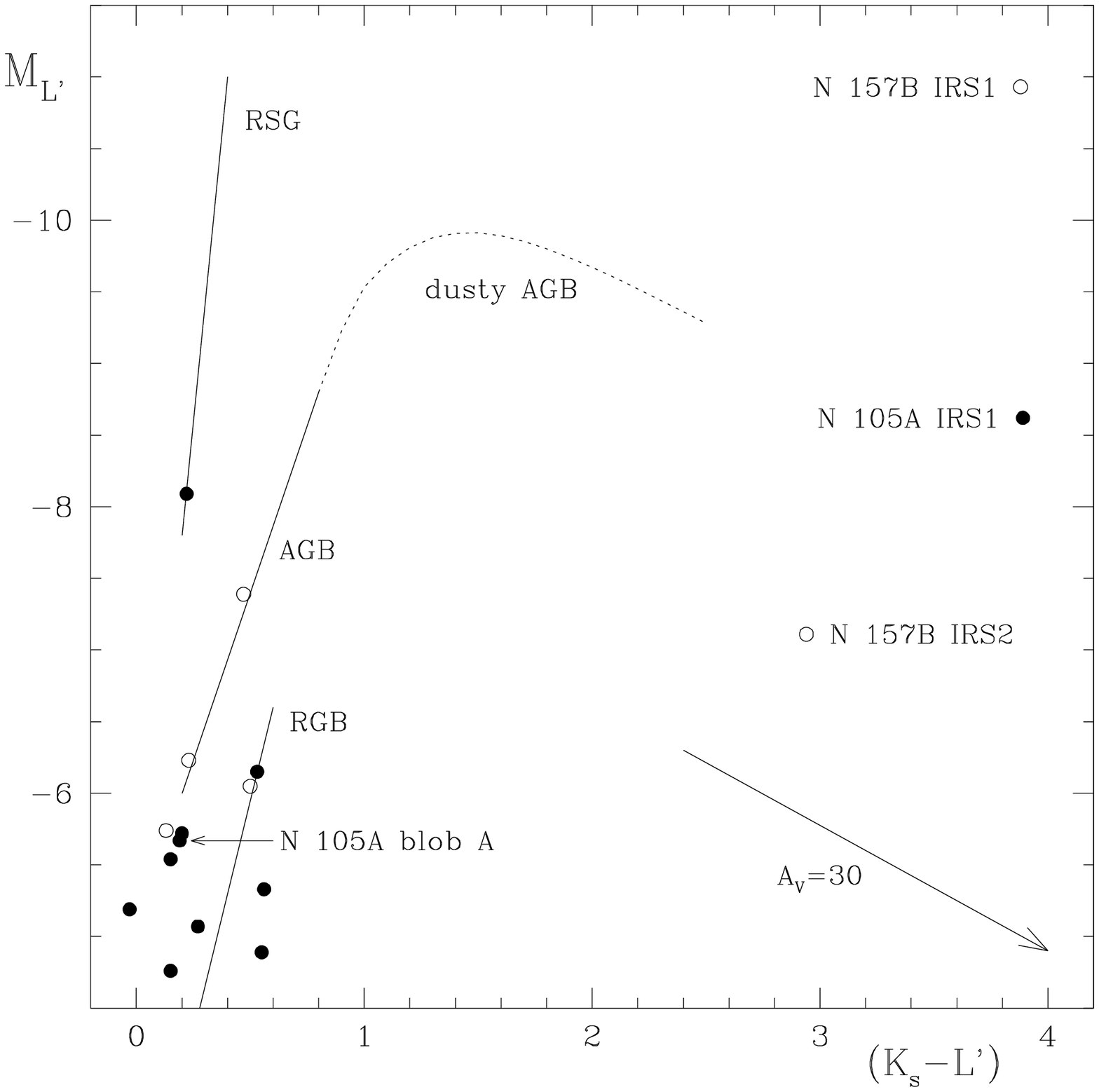}
\caption[]{Absolute L$^\prime$-band magnitudes versus K$_{\rm s}$--L$^\prime$
colours for objects near N\,157B (open circles) and N\,105A (filled circles). 
Approximate empirical sequences from \citet{vanloon05a} are overplotted for the 
red giant branch (RGB), red supergiants (RSG) and asymptotic giant branch (AGB, 
with mass-loss evolution); early type main-sequence stars have 
K$_{\rm s}-$L$^\prime \sim 0$. The effect of $A_{V} = 30$\,mag extinction is 
also shown. Only three of the sources in this diagram are particularly red.}
\label{nir}
\end{figure}

The L$^\prime$-band acquisition images of N\,157B and N\,105A allow us to 
identify a number of red sources other than the two spectroscopic targets. 
Aperture photometry was performed on all sources in each image. Very red objects
(typically $K_{\rm s}-L^\prime>1$\,mag) are referred to as ``IRS'', and these 
are numbered on the basis of the $K_{\rm s}-L^\prime$ colour and L$^\prime$-band
brightness. These IRS objects have significant excess emission at 3\,$\mu$m 
(Fig.\,\ref{nir}) and are not simply reddened objects behind the molecular 
clouds. The other point sources are referred to as ``S'', and these are numbered
with increasing RA. Two other sources in N\,105A seem to be diffuse in nature, 
at least in the L$^\prime$-band, so they are named blob A and B (see discussion
below). The identification, 2MASS positions and magnitudes and L$^\prime$-band 
magnitudes are listed on Table\,\ref{nir_t}. In the next subsections these 
sources and the IR spectra are described in detail.

\subsubsection{Infrared sources in N\,157B}
\label{n157spec}

MSX\,LMC\,888 \citep[=IRAS\,05381$-$6912,][]{egan01} is a bright
mid-IR point source in the 30\,Dor region, situated within $1^\prime$ from 
N\,157B (usually identified as SNR\,0538$-$69.1, although \citet{chu04}
question the nature of the nebula as a supernova remnant). Close to N\,157B 
there is a small molecular cloud, JGB\,30\,Dor-22, with a diameter of 6.8\,pc 
identified from CO observations \citep{johansson98}. The cold emission from 
the molecular cloud appears spatially resolved on MSX images, whilst the 
bright mid-IR source stands out through its warmer unresolved emission.

In our L$^\prime$-band image (Fig.\,\ref{vlt_l}), the MSX source is easily 
identified as a bright IR star, that we call N\,157B\,IRS1. There is at least 
one other star nearby also with extreme $K_{\rm s}-L^\prime$ colour, 
N\,157B\,IRS2. IRS2 is seen in the direction of the molecular cloud and 
may be located behind it, but its very extreme $K_{\rm s}-L^\prime$ colour 
suggests it could also be a young object embedded in the cloud, warming the 
surrounding dust. IRS2 is considerably fainter than IRS1. N\,157B\,S1 and S4 
have $(J-K_{\rm s}) >$\,1.3\,mag but $(K_{\rm s}-L^\prime)<$\,0.5\,mag and are 
likely suffering from extinction by dust within the molecular cloud.

\begin{figure}
\includegraphics[width=84mm]{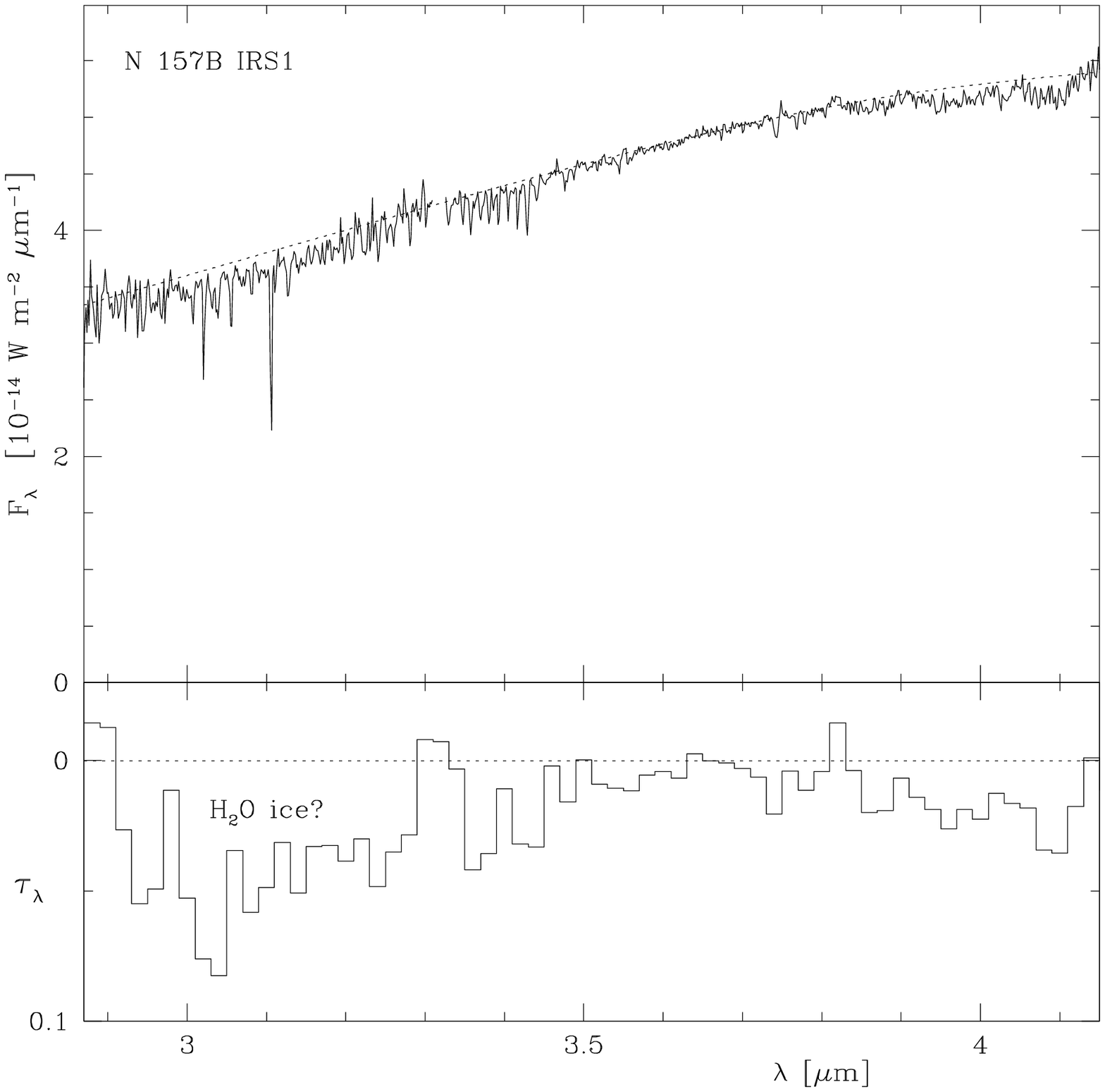}
\caption[]{Spectrum of the massive YSO candidate N\,157B\,IRS1 in the 30\,Dor 
region. The strong 3.32\,$\mu$m telluric methane feature has been removed. The 
spectrum shows no H recombination emission lines. The lower panel shows the 
optical depth with respect to the continuum (dotted line in the top panel); 
there might be a hint of the broad water ice feature at 3.1\,$\mu$m.}
\label{msx888_lspec}
\end{figure}

The spectrum of N\,157B\,IRS1 (Fig.\,\ref{msx888_lspec}) seems at first glance 
featureless. It does not resemble spectra of evolved stars, e.g. a heavily 
reddened, late-M type giant \citep[cf.][]{matsuura05,vanloon06}. There is also 
no evidence for H recombination emission lines arising from an ionised region 
(Sect.\,\ref{n105spec}). When the spectrum is plotted as optical depth with
respect to a pseudo-continuum (a low-order polynomial fit) there seems to be a 
hint of the broad water ice feature at 3.1\,$\mu$m, similar to that in 
IRAS\,05328$-$6827 \citep{vanloon05b} also in the LMC. The evidence is, however,
not conclusive with a column density at most $N({\rm H}_2{\rm O})<10^{17}$ 
cm$^{-2}$. It is unclear what are the narrow absorption features blueward of 
3.5\,$\mu$m.

\subsubsection{Infrared sources in N\,105A}
\label{n105spec}

MSX\,LMC\,80 \citep[=IRAS\,05101$-$6855,][]{egan01} is a bright IR source
in the H\,{\sc ii} region LHA\,120-N\,105A. It is associated with an IR object
that was suggested by \citet*{epchtein84} to be a ``protostar''. This object, 
which we shall refer to as N\,105A\,IRS1, is very bright in the L$^\prime$-band 
(Fig.\,\ref{vlt_l}) and with $K_{\rm s}-L^\prime=3.9$ mag it is extremely red 
(Table\,\ref{nir_t}). It is not detected in the J and H-band of 2MASS, so it is
probably heavily extincted.

LHA\,120-N\,105A\,blob is associated with a small but extended nebulosity at a 
projected distance of $14^{\prime\prime}$ from IRS1 and $10^{\prime\prime}$ from
S8, the WR star Br\,16a \citep{dopita94}. We identify a bright core, blob\,A, 
surrounded by patchy emission visible in the L$^\prime$-band image 
(Fig.\,\ref{vlt_l}). This diffuse core has a point-source 2MASS counterpart, but
the 2MASS images seem to show some diffuse emission too. We have assigned 
an identifier to one of these patches, blob\,B, because it happened to fall in 
the slit of the spectrograph and produce a clear signal (see below). Continuum 
emission was detected at 6.6 GHz \citep{ellingsen94}, centred on IRS1 and 
blob\,A, connecting the two sources. It thus appears that both these sources 
are embedded in an ionised environment (see below), probably created by IRS1 
itself or a source within blob\,A. The $(K_{\rm s}-L^\prime)$ colour of the 
bright knot blob\,A is not particularly red (Fig.\,\ref{nir}) even though 
$(J-K_{\rm s})$\,=\,1.51\,mag suggests some extinction.

IRS2 is a very faint source located within $\sim0.5^{\prime\prime}$ from the 
positions of the water \citep{lazendic02} and OH masers \citep{brooks97}. It was
not detected in the 2MASS survey, but appears as a very faint star in our 
L$^\prime$-band image (Fig.\,\ref{vlt_l}). Given the low density of stars in the
image the association with the masers is suggestive. This object could be an
embedded YSO. The methanol maser detected by \citet{sinclair92} unfortunately 
falls just outside the SE corner of the field of the L$^\prime$-band image 
\citep[see also][]{ellingsen94}.

Several more sources can be identified from Fig.\,\ref{vlt_l}. They all appear 
stellar in nature and have $(K_{\rm s}-L^\prime)<$\,0.6\,mag. S6 and S9 have
$(J-K_{\rm s})>$\,1.2\,mag thus they could be situated behind the molecular 
cloud. The remaining sources have unremarkable near-IR colours.

\begin{figure}
\includegraphics[width=84mm]{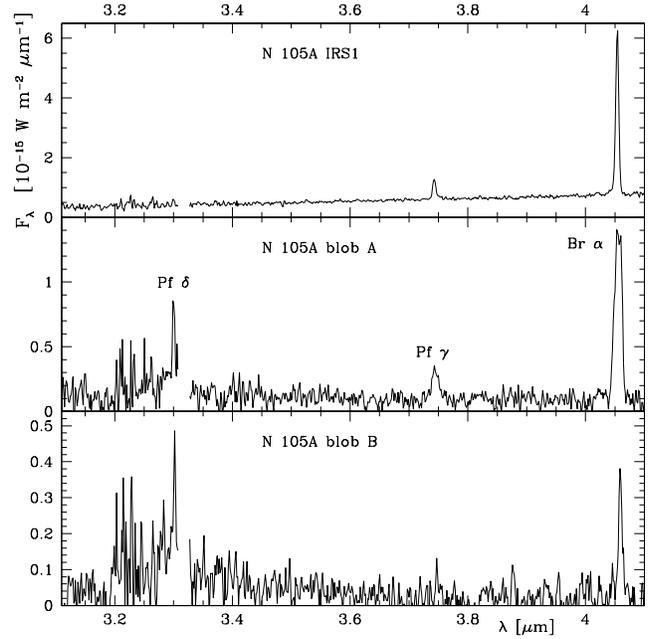}
\caption[]{Spectra of the massive YSO candidate, N\,105A\,IRS1, and the blobs 
N\,105A\,blob\,A \& B. The strong 3.32\,$\mu$m telluric methane feature has been
removed. Besides line emission from Br$\alpha$ and Pf$\gamma$ and $\delta$, IRS1
has a spatially unresolved (very) red continuum whilst blob\,A \& B display 
broad 3.3 $\mu$m emission attributed to PAHs.}
\label{n105_lspec}
\end{figure}

\subsubsection{Hydrogen line emission and outflows in N\,105A}

The spectrum of  N\,105A\,IRS1 shows a very red continuum, very likely free-free
emission (see Sect.\,\ref{seds}), underlying H recombination line emission 
(Fig.\,\ref{n105_lspec}). There is no evidence in the spectrum for broad ice
features. The Br$\alpha$ line is about eight times as bright as the continuum, 
but the Pf$\delta$ line cannot be positively detected. This suggests a high 
dust column density in front of the region in which the line emission arises. 
The Br$\alpha$ line emission is blended at this resolution with emission from 
He\,{\sc i} at 4.0378, 4.0410 and 4.049\,$\mu$m; this emission is very weak when
compared to Br$\alpha$ \citep*{drew93} so it has a negligible effect on the
measured fluxes. We have measured the line fluxes of Br$\alpha$ and Pf$\gamma$;
the Br$\alpha$ to Pf$\gamma$ flux ratio is $\sim$\,9. This ratio can be compared
to the prediction from recombination theory. We consider case B treatment 
\citep[e.g.,][]{hummer87} and adopt $T_{\rm e}=7500$\,K and 
$n_{\rm e}=1000$\,cm$^{-3}$ --- the emission coefficients are not very 
sensitive to temperature or density. Using the \citet{hummer87} computed values,
the theoretical flux ratio is predicted to be $\sim$\,7.5. If we adopt a simple 
power law for the extinction as a function of wavelength \citep{mathis90}, we 
estimate that the Br$\alpha$ to Pf$\gamma$ flux ratio for IRS1 is consistent 
with case B scenario affected by $A_{\rm V} \sim 40$\,mag. This agrees very well
with the reddening needed to fit the object's Spectral Energy Distribution 
(SED, Sect.\,\ref{seds}).

\begin{figure}
\includegraphics[width=84mm]{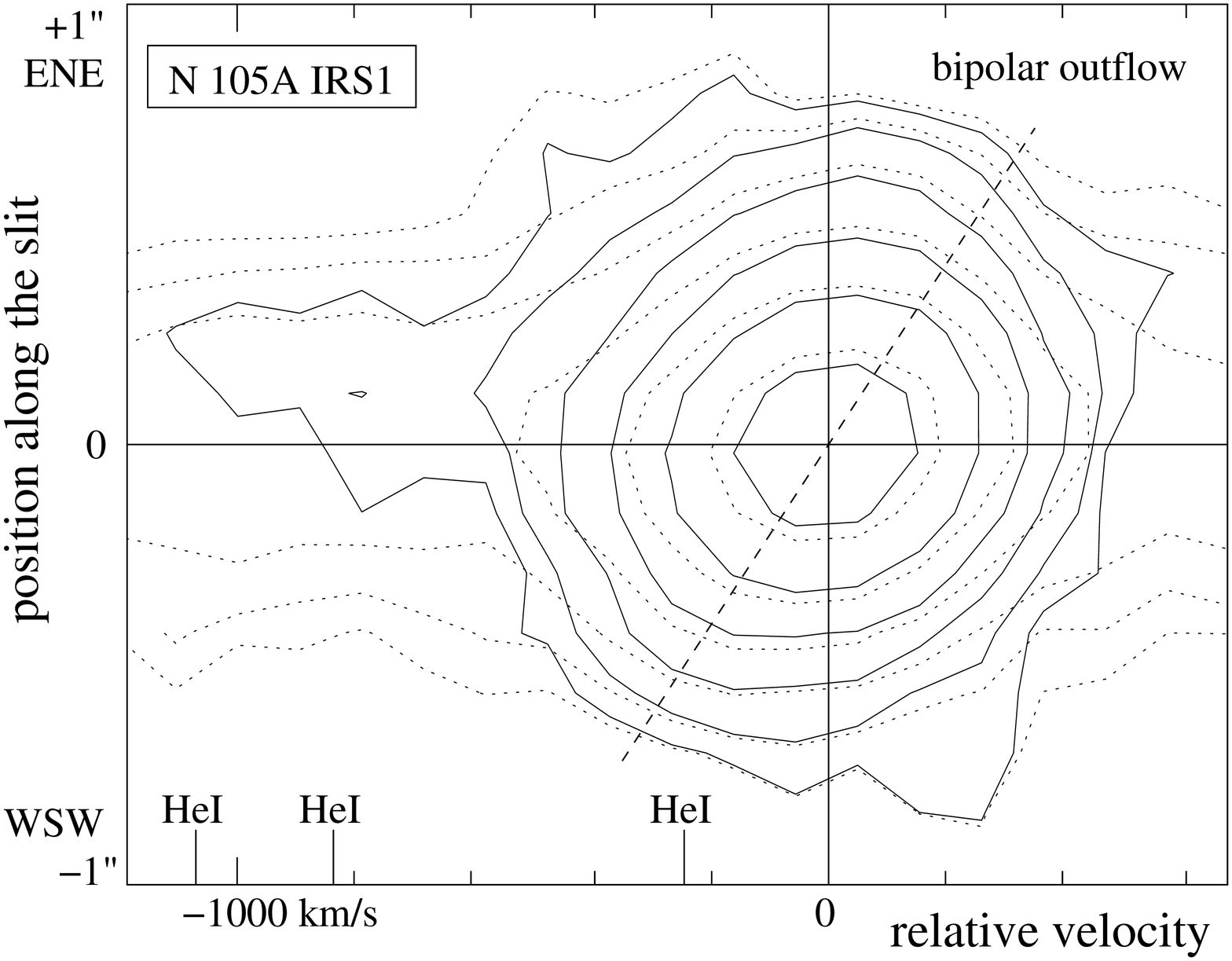}
\caption[]{Position-velocity diagram of the massive YSO, N\,105A\,IRS1, showing 
emission in the Br$\alpha$ line before (dotted) and after (solid) subtracting 
the continuum emission. The contour levels range from $6\times10^{-6}$ to 
$1.9\times10^{-4}$\,W\,m$^{-2}$\,$\mu$m$^{-1}$\,sr$^{-1}$ in increments of a 
factor 2. The blueward wing of the Br$\alpha$ line profile is affected by blends
with very weak He\,{\sc i} emission, approximately at $-1070$, 835 and 
$-245$\,km\,s$^{-1}$, responsible for the emission tail towards large 
blueshifted velocities. There is a clear asymmetry in the Br$\alpha$ profile 
(highlighted by the dashed line), evidence for a bipolar outflow.}
\label{proto_spec}
\end{figure}

The Br$\alpha$ line is marginally resolved in IRS1, both spatially and 
kinematically. To study the spatial distribution of the Br$\alpha$ emission we 
plot a position-velocity diagram, in which one axis is the dispersion direction 
and the perpendicular axis is the spatial direction along the slit 
(Fig.\,\ref{proto_spec}). A spatial profile for the continuum emission is 
obtained by averaging many columns on the array at either side of the emission
line. After subtracting this continuum (solid lines in Fig.\,\ref{proto_spec}),
the line profile has broad wings and shows a clear asymmetry with emission in 
the WSW moving towards us and emission in the ENE moving away. This is strong 
evidence for a bipolar outflow originating near the centre of IRS1, at a 
projected velocity of $v_{\rm bipolar}\sim100 - 200$\,km\,s$^{-1}$ at a few 
tenths of an arcsecond ($\sim0.1$ pc) at either side of the star. 

Blobs\,A and B also show emission in the H recombination lines but in this
case also Pf$\delta$ is detected (Fig.\,\ref{n105_lspec}), indicating that 
extinction by dust is less severe than towards IRS1. Continuum emission in
this source is extremely faint, if at all present. Both components show evidence
for broad emission around 3.3\,$\mu$m underlying the Pf$\delta$ line. The shape
of this emission is consistent with the 3.28 $\mu$m unidentified IR feature that
is usually attributed to Polycyclic Aromatic Hydrocarbons 
\citep[PAHs,][]{allamandola85}. The Br$\alpha$ to Pf$\gamma$ flux ratio for 
blob\,A is $\sim$\,5, which might suggest that Br$\alpha$ is optically thick 
\citep{drew93}. 

\begin{figure}
\includegraphics[width=84mm]{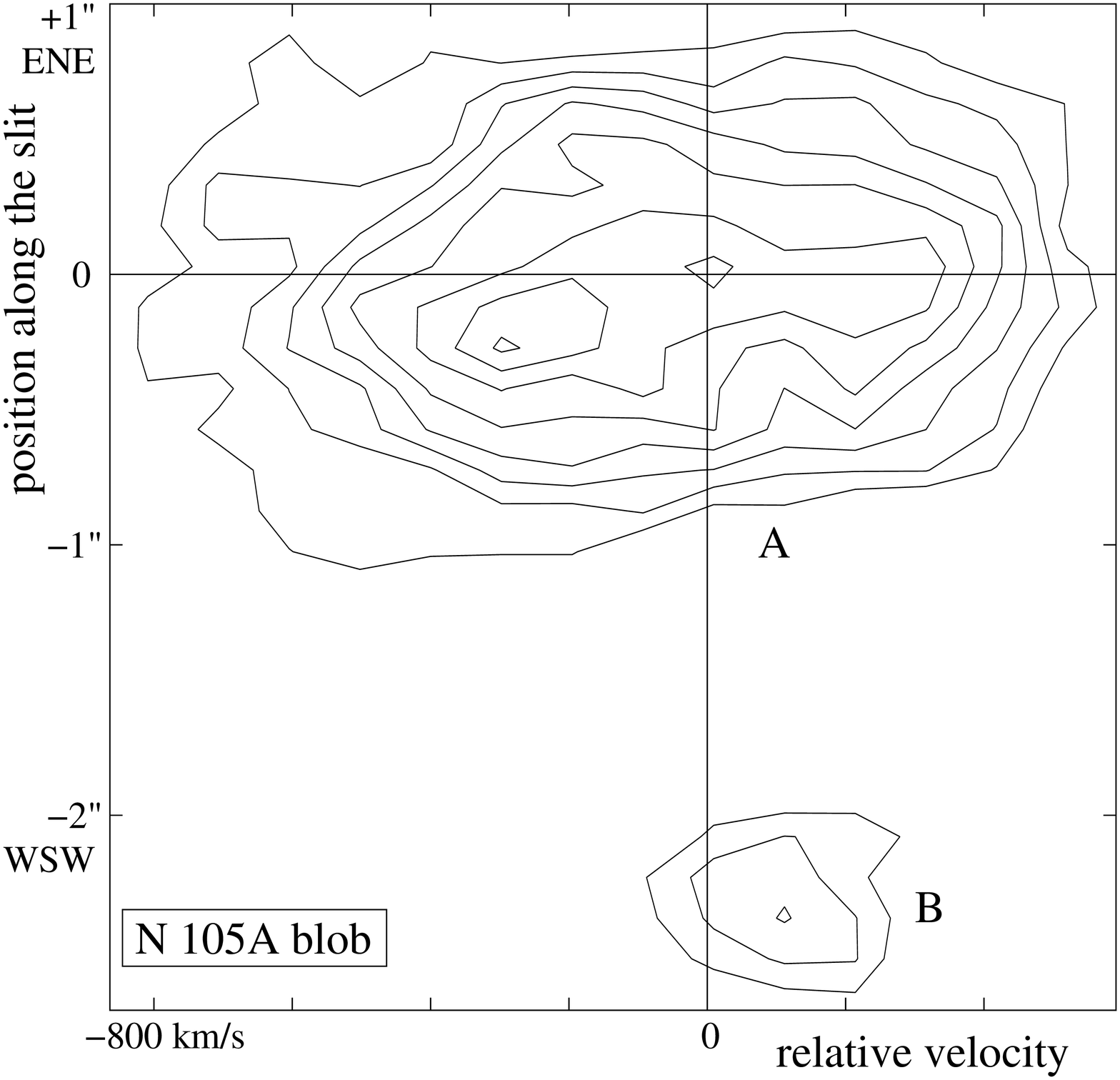}
\caption[]{Position-velocity diagram of the N\,105A\,blob\,A+B nebulosity, 
showing emission in the Br$\alpha$ line. The contour levels range from 
$1.1\times10^{-5}$ to $3.5\times10^{-5}$ W m$^{-2}$ $\mu$m$^{-1}$\,sr$^{-1}$ in 
increments of $3.5\times10^{-6}$ W m$^{-2}$ $\mu$m$^{-1}$\,sr$^{-1}$. Blob\,A 
resembles an expanding, asymmetric shell whereas blob\,B is a compact, 
kinematically cold cloud.}
\label{proto_spec2}
\end{figure}

The spectrum of blob\,A seems broader than IRS1 and exhibits a double-peaked 
line profile (Fig.\,\ref{n105_lspec}), suggestive of an outflow also emanating 
from this nebular object. The position-velocity diagram of the Br$\alpha$ line 
(Fig.\,\ref{proto_spec2}) shows a kinematically broad emission suggesting a 
compact outflow at a projected velocity of 
$v_{\rm outflow}\sim\pm300$\,km\,s$^{-1}$. The spectrum shows departures from 
spherical symmetry: the brightest emission, at $\sim0.2^{\prime\prime}$ to the 
WSW of the core of blob\,A, is associated with gas that is preferentially moving
towards us. Blob\,B, on the other hand, is kinematically cold and shows no signs
of outflow. This seems to suggest that there might be a source powering an 
outflow in blob\,A, even if there is no evidence of free-free continuum from an 
ionised medium.

\subsection{Spectral Energy Distributions of N\,157B\,IRS1 and N\,105A\,IRS1}
\label{seds}

We performed aperture photometry in the IRAC images for the objects identified
in the L$^\prime$-band images, obtaining both fluxes and magnitudes
\citep{reach05}. The 3.6\,$\mu$m magnitudes agree quite well for most sources in
N\,105A and N\,157B, except for the objects in areas where the ``nebulosity'' 
appears denser (Figs.\,\ref{30dor_8m} and \ref{mir}). In these areas the 
background is not correctly estimated, and measured fluxes are depressed. 
Nevertheless, the colours of these objects can still provide another hint on 
their nature. We briefly discuss here only the objects with measurements in the 
4 IRAC bands: N\,105A\,IRS1, IRS2, blob\,A+B and S2, and N\,157B\,IRS1 and IRS2.

\citet[][ and references therein]{jones05} defines the locus for red giants, 
Class\,II YSOs (objects with disks), Class\,I YSOs (embedded objects) and 
H\,{\sc ii} regions. N\,105A\,S2 has colours consistent with it 
being a red giant. N\,105A\,IRS2, N\,157B\,IRS2 and N\,105A\,blob\,A+B have 
colours consistent with H\,{\sc ii} regions, i.e.\ regions with ionised gas 
and less dust column density. N\,105A\,IRS1 has colours consistent with a very 
reddened compact H\,{\sc ii} region, in agreement with the presence of H 
recombination lines in its spectrum, reddening determination and the findings of
the SED analysis (see below). N\,157B\,IRS1 has colours consistent with an 
embedded (Class\,I) YSO.

\begin{table*}
\caption[]{Mid-IR photometry (in Jy) of the embedded YSOs N\,157B\,IRS1
and N\,105A\,IRS1: IRAC 3.6, 4.5, 5.8 and 8.0\,$\mu$m, MSX 8.28, 12.1, 14.7 and 
21.3\,$\mu$m, IRAS 12, 25, 60 and 100\,$\mu$m. IRAC flux densities are 
uncertain by up to 10 per cent.}
\label{mir_t}
\begin{tabular}{l@{\hspace{3.1mm}}c@{\hspace{3.1mm}}c@{\hspace{3.1mm}}c@{\hspace{3.1mm}}c@{\hspace{3.1mm}}c@{\hspace{3.1mm}}c@{\hspace{3.1mm}}c@{\hspace{3.1mm}}c@{\hspace{3.1mm}}c@{\hspace{3.1mm}}c@{\hspace{3.1mm}}c@{\hspace{3.1mm}}c}
\hline
Name &\multicolumn{4}{c}{IRAC} &\multicolumn{4}{c}{MSX}&\multicolumn{4}{c}{IRAS} \\
     &3.6\,$\mu$m  &4.5\,$\mu$m &5.8\,$\mu$m&8.0\,$\mu$m &8.28\,$\mu$m &12.1\,$\mu$m &14.7\,$\mu$m &21.3\,$\mu$m &12\,$\mu$m   &25\,$\mu$m   &60\,$\mu$m  &100\,$\mu$m  \\
\hline
N\,157B\,IRS1 &0.173& 0.313&0.483&0.635&0.96$\pm$0.04&1.62$\pm$0.09&1.89$\pm$0.12&4.2$\pm$0.3	   &4.0$\pm$0.3&25$\pm$3&170$\pm$20&140$\pm$40\\
N\,105A\,IRS1 &0.028& 0.063&0.181&0.473&1.72$\pm$0.07&4.70$\pm$0.24&6.61$\pm$0.40&\llap{2}0.7$\pm$1.2 &8.0$\pm$0.3&56$\pm$2&310$\pm$10&400$\pm$20\\
\hline
\end{tabular}
\end{table*}

\begin{table*}
\caption[]{Input parameters for, and results from the modelling with the {\sc
dusty} code of the spectral energy distributions of the embedded YSOs 
N\,157B\,IRS1 and N\,105A\,IRS1. Both make use of a standard MRN grain
size distribution (see text).}
\label{sed_t}

\begin{tabular}{lllcccccc}
\hline
Name                        &central source              &dust type                        &$T_{\rm in}$ & $\rho_{\rm in}$    & $r_{\rm in}$&$r_{\rm out}$ & $A_{\rm V}$ &$L$ \\
                            &                            &                                 & (K)         &  (g\,cm$^{-3}$)    & (AU)        & (pc)         & (mag)       &(L$_\odot$)\\
\hline
N\,157B\,IRS1               &25\,000 K blackbody         &silicate                         &730          &$8 \times 10^{-19}$ & 350	    &1.7	   &23 	 &$7.8 \times 10^4$\\
N\,105A\,IRS1               &free-free (911 \AA...10 cm) &0.7 silicate + 0.3 carbo\rlap{n} &350          &$2 \times 10^{-19}$ &\llap{2}500  &1.2	   &48 	 &$2.5 \times 10^5$\\
\hline
\end{tabular}
\end{table*}

\begin{figure}
\includegraphics[width=84mm]{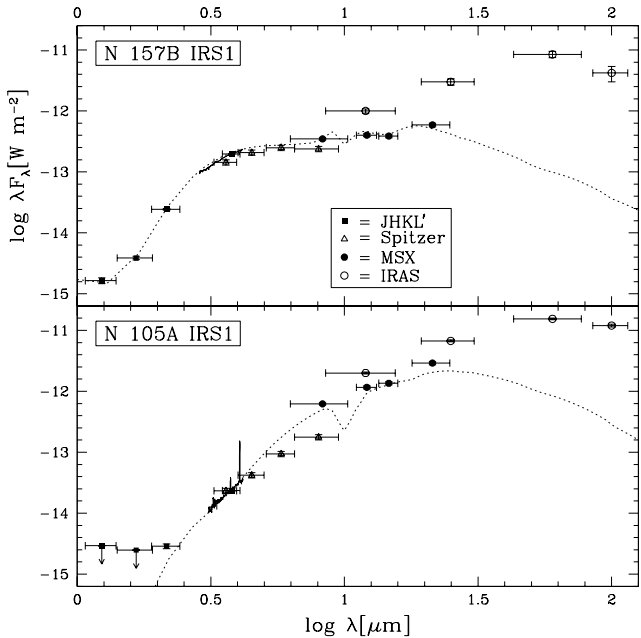}
\caption[]{Spectral energy distributions of N\,157B\,IRS1 (top) and 
N\,105A\,IRS1 (bottom). The IRAS data have poor spatial resolution and probably 
include emission from cold dust surrounding the YSO candidate. Fits obtained 
with the {\sc dusty} code are overplotted as dotted curves (see text).}
\label{sed_f}
\end{figure}

The spectral energy distributions (SEDs) of N\,105A\,IRS1 and N\,157B\,IRS1
were reproduced with the dust radiative transfer model {\sc dusty}
\citep{ivezic99}, as an attempt to derive approximate quantitative information 
about these YSOs. The fits are shown along with the observed SEDs in 
Fig.\,\ref{sed_f}, and the main fit parameters are summarised in 
Table\,\ref{sed_t}. In both cases we could obtain a reasonable fit to the SED 
with a radial dust density profile 
$\rho(r) = \rho_{\rm in} (r/r_{\rm in})^{-1.5}$ (steady accretion), and a 
standard MRN grain size distribution \citep{mathis77}. The dust envelope was 
assumed to be spherically symmetric; this is probably not strictly true. The
bolometric luminosity was derived from scaling the model SED to fit the observed
SED and using a value for the distance to the LMC of 50\,kpc.

The main difference between the two objects is in the nature of the central
source. In the case of N\,157B\,IRS1 we place a black body of 25,000\,K in the
centre of the dust envelope. Such hot star is required to reproduce the SED
throughout the $JHK$ bands where the star becomes noticeable; at longer
wavelengths (including the 3$-$4\,$\mu$m band) the SED is dominated by dust
emission. Based on the SED analysis alone, we cannot entirely rule out the 
possibility that this is an evolved dust-enshrouded red supergiant of around 
10$-$15\,M$_\odot$. However, this would imply a rather long ($\sim20$\,Myr) 
time-gap between the formation of this massive star and the star formation that
is currently taking place in the 30\,Dor region --- \citet{walborn97} identified
5 different stellar populations in the 30\,Dor nebula, with ages in the range 
$<1$ and 10\,Myr. A chance coincidence of such rare object with N\,157B is 
extremely unlikely. Furthermore the cautious identification of water ice 
(Sect.\,\ref{n157spec}) argues against it. 

In the case of N\,105A\,IRS1, the 3$-$4\,$\mu$m spectrum clearly indicates the
presence of an ionised region inside of the dust envelope. We thus represented 
the central source by a free-free emission object. There can be little doubt 
that this is a young object and the surrounding dust is not produced by the 
central star, and we therefore used a mixture of amorphous oxygen-rich silicates
\citep*{ossenkopf92} and carbon \citep{hanner88}. The SED is rather well 
reproduced, especially throughout the 3$-$4\,$\mu$m band, and the optical
extinction used in the model is consistent with that estimated from the 
hydrogen emission line ratios (see Sect.\,\ref{n105spec}). The dust envelope is 
cold, again consistent with a cavity created by the emerging H\,{\sc ii} region 
powered by a massive YSO. The bolometric luminosity of N\,105A\,IRS1 suggests a 
central star of spectral type O5 and about 43\,M$_\odot$ \citep{hanson97}.

\subsection{H$\alpha$\ map of N\,113}
\label{halpha}

\begin{figure*}
\includegraphics[width=84mm]{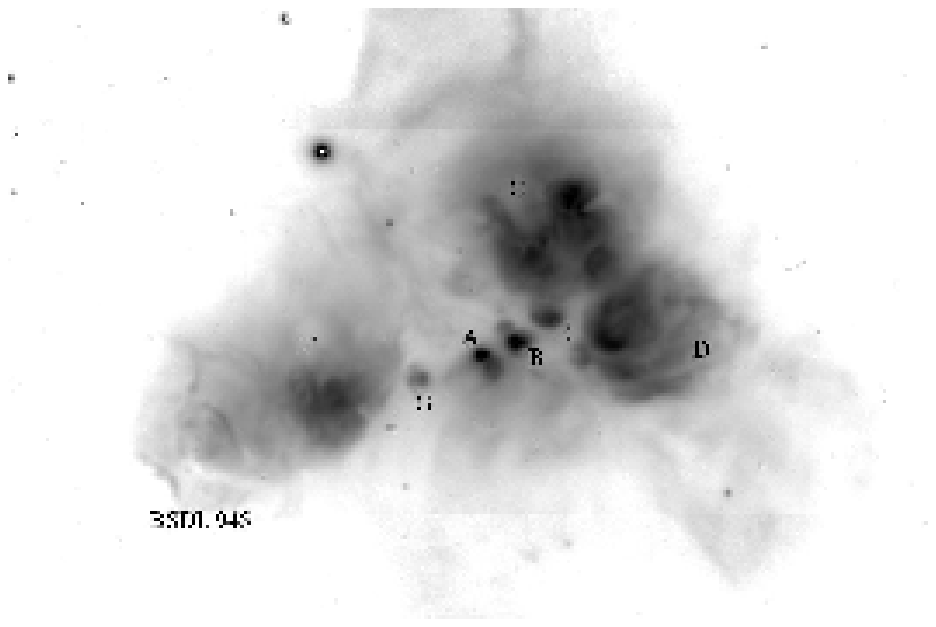}
\includegraphics[width=84mm]{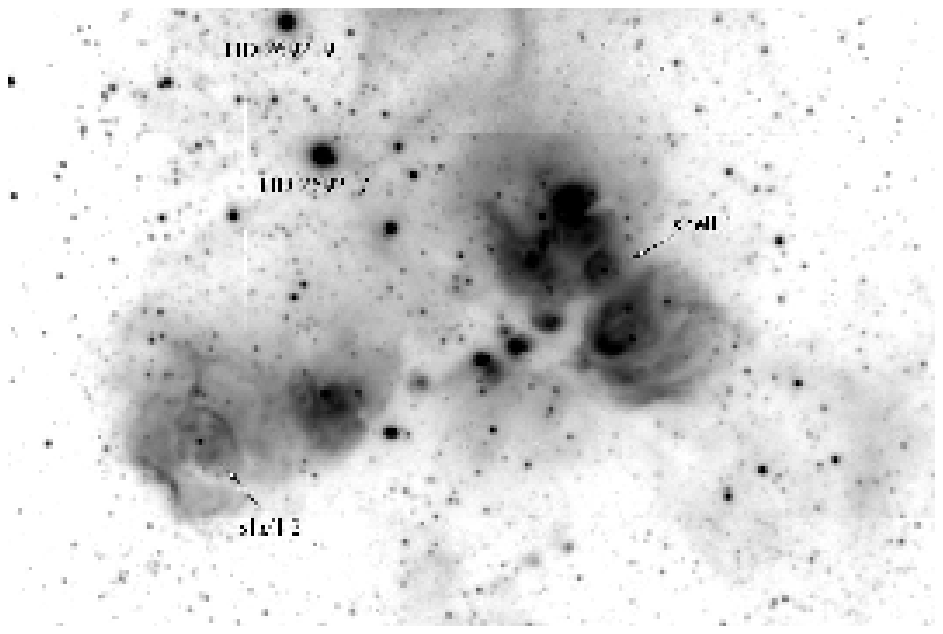}
\includegraphics[width=84mm]{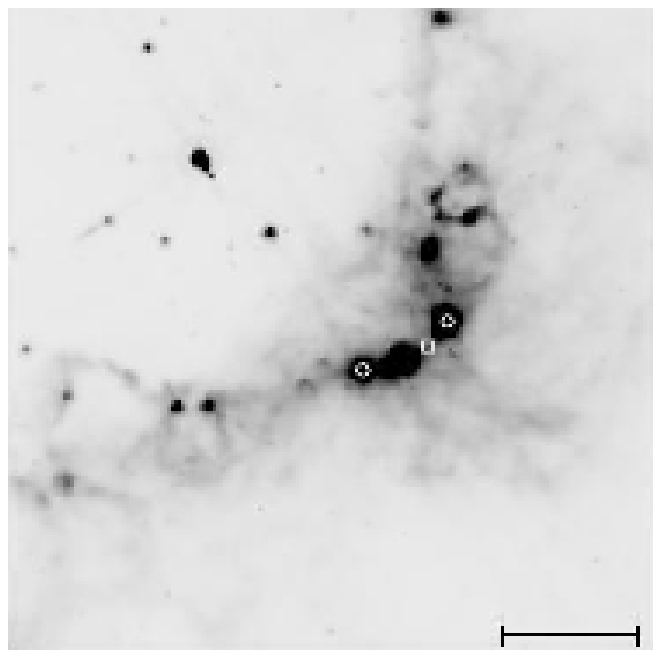}
\includegraphics[width=84mm]{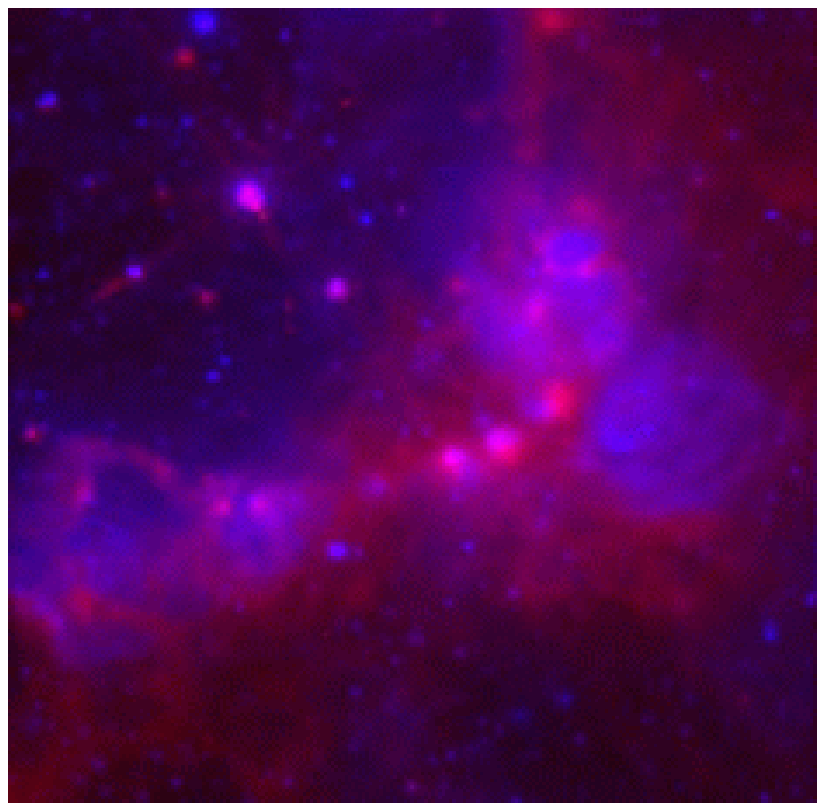}
\caption[]{Optical and IR images of N\,113. The H$\alpha$ emission (top left) 
and pseudo-continuum (top right) images cover approximately 
4.4$\arcmin \times$\,6.7$\arcmin$ centered at $\sim 05^h13^m25^s,
-69\degr22\arcmin26\arcsec$. The greyscale is logarithmic between 10$^4$ and 
3$\times 10^5$ Rayleigh for the H$\alpha$ emission image and between 200 and 
3000\,R for the continuum image. The most important gas structures and stars are
identified and labelled. The 8$\mu$m IRAC/Spitzer image is shown at the bottom 
left (4.7$\arcmin \times$\,4.7$\arcmin$, centered at $\sim 05^h13^m27^s,
-69\degr22\arcmin28\arcsec$; the scale indicates 1$\arcmin$; the greyscale is 
linear between $-$5 and 100\,MJy\,sr$^{-1}$). Maser sources
\citep[circles;][]{lazendic02,brooks97} and a molecular core 
\citep[square;][]{wong06} are identified. At the bottom right is a false colour 
image where H$\alpha$ emission and stars appear in blue while red is molecular 
cloud material bright in the 8\,$\mu$m image. No attempt was made to remove 
artifacts from the IR image, for instance the ghost impressions caused by the 
brightest stars. In all images North is to the top and East to the left.}
\label{n113}
\end{figure*}

The H$\alpha$ images of N\,113 (Fig.\,\ref{n113}) show a wealth of detail. 
The line emission is so strong that it is still apparent in the 
pseudo-continuum image (though at a much fainter level). The main extended 
structures are three regions of H$\alpha$ emission to the east, north and west 
of the centre of the H\,{\sc ii} region: N\,113F, N\,113C and N\,113D, 
respectively. This is emphasised by an obscuring lane running across the 
H\,{\sc ii} region, onto which are projected (from east to west) N\,113A, 
N\,113B and N\,113E: ``a chain of three small intense knots of nebulosity which,
together, make up NGC\,1877'' \citep{henize56}. There is a fourth such knot ---
which we name N\,113G --- in between N\,113A and N\,113F. Although fainter, it 
stands out well against the dark dust lane.

N\,113C and the lower surface brightness nebula BSDL\,945 \citep{bica99}
to the east of N\,113F each contain a particularly striking example of a shell
with a central star (indicated in Fig.\,\ref{n113} as ``shell 1'' and ``shell
 2''). These shells are likely to be stellar wind-blown bubbles with a dynamical
timescale of only a few $\sim10^3$\,yr (for a wind speed of order 
$10^3$\,km\,s$^{-1}$). The central star of shell 1 can be identified with a 
B0$-$0.5\,III star \citep{wilcots94}, but the central star of the larger shell 2
remains anonymous.

Several OB stars in N\,113 have been described in the literature. Near the
edge of the H$\alpha$ emission, the evolved B2[e] supergiant HD\,269217 (=
Hen\,S\,89, R\,82, IRAS\,05136$-$6925, MSX\,LMC\,216) has enjoyed considerable
attention. It was first mentioned in \citet{merrill33} as a B star with strong 
Balmer line emission. Descended from a 30\,M$_\odot$ main sequence star, it has
been associated with circumstellar dust and found to move at a heliocentric 
speed of $v_\star\sim240$\,km\,s$^{-1}$ \citep{zickgraf86}. Another, little 
studied emission-line star HD\,269219 (= Hen\,S\,90) is found further north and 
away from the H\,{\sc ii} region. Closer to the core of the H\,{\sc ii} complex,
HV\,2377 is an M-type suspected supergiant. Both around and embedded within 
N\,113 are a number of O9$-$B0.5 stars --- either on the main sequence or 
slightly evolved \citep{brunet75,wilcots94}.

\section{Discussion}

\subsection{Triggered star formation in the mini-starburst 30\,Doradus and 
N\,157B}

The central area of the 30\,Dor nebula (N\,157A) exhibits a complex star 
formation history. As already mentioned, \citet{walborn97} disentangled five 
distinct stellar populations in the central area of 30\,Dor. The population 
located mainly in the molecular filaments to the west and northeast of R\,136 
shows spectroscopic evidence of extreme youth ($<1$ Myr). These infant objects 
are spatially related with bright, compact IR sources within a complex nebular 
and dust morphology \citep{hunter95}: pillars of molecular gas and dust and dark
globules \citep*{walborn99,walborn02}, which are being photo-evaporated by the 
intense radiation from the massive compact cluster R\,136 \citep{hunter95}.
Together with the maser sources, this constitutes strong evidence that star
formation is ongoing in the nebula.

\begin{figure}
\includegraphics[width=84mm]{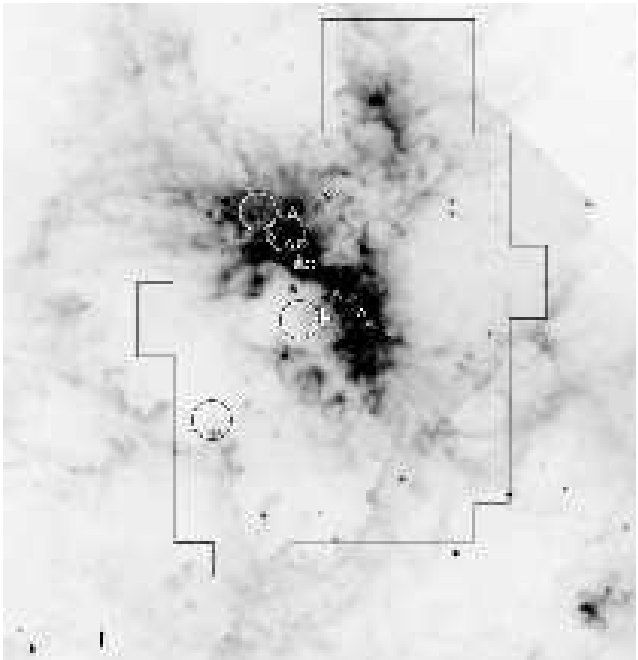}
\caption[]{IRAC/Spitzer 8.0\,$\mu$m image of 30\,Doradus (N\,157A), with N\,157B
at the bottom right of the image (North to the top, East to the left; the scale
indicates 1\,arcmin). The greyscale of the image is linear between $-$5 and 
150\,MJy\,sr$^{-1}$. Large circles are the maser detections presented in this 
paper (the size of the circle represents the approximate positional accuracy) 
while small circles are maser positions from \citet[][ water maser]{lazendic02} 
and \citet[][ OH maser located slightly to the south and west of the water maser
position]{brogan04}. Triangles are embedded objects (see text) from 
\citet{brandner01} and \citet{rubio98}. The large outline represents the area 
covered by our maser survey. The cross indicates the position of R\,136 ($05^h
38^m42^s,-69\degr06\arcmin03\arcsec$).}
\label{30dor_8m}
\end{figure}

Fig.\,\ref{30dor_8m} shows the 8.0\,$\mu$m IRAC/Spitzer image of 30\,Dor. At 
this wavelength, the extended emission is dominated by hot dust and PAH 
emission. This figure also shows the location of young, embedded IR sources 
($J-H$ and $H-K$ $\ge$\,1.5\,mag) identified in the literature 
\citep{rubio98,brandner01,maercker05} as well as the maser sources. The water 
masers in 30\,Dor are located in similar structures as the IR sources, 
surrounding R\,136, suggesting that {\it current} star formation (not just 
{\it recent} star formation) is also concentrated near the interfaces of the 
molecular cloud complex within the influence sphere of the previous generation 
of massive O and B-type stars. The large velocity difference between maser 
source components (Sect.\,\ref{maser_30dor}) implies location in distinct 
structures in the ISM. As proposed by \citet{vanloon01}, the highly supersonic 
velocities of some of the sources and their location near the rim of large gas 
superbubbles \citep[e.g.,][]{wang91} strongly suggest that the masers occur at 
the collision fronts of rapidly expanding bubbles of ionised gas with the 
surrounding dense neutral material. If the masers trace the velocities of 
protostars within these environments, then this constitutes evidence that star 
formation was triggered at different locations by the massive stars' feedback. 

Our maser observations survey the central part of the 30\,Dor nebula, covering
the immediate neighbourhood of R\,136 and the densest regions of the nebula.
Even though the survey area is much larger than previous observations, we only 
identified one new maser source. We can compare our survey results to the 
galactic water maser luminosity distribution from \citet{valdettaro01}. Assuming
a median distance to high-mass star forming regions in nearby spiral arms of a 
few kpc, we would expect the bulk of the water masers in 30\,Dor to have 
integrated fluxes of order 0.1$-$1 Jy\,km\,s$^{-1}$, with a steep decline at the
higher end. This suggests that, although close, our survey is not yet deep 
enough to reveal the bulk of the water maser population in 30\,Dor. Nonetheless,
as all detected water masers are located at or near compressed interface 
regions, our survey provides evidence that most of the on-going massive star 
formation occurs in the presence of feedback.

The situation is less clear for the molecular material in N\,157B, still in the
30\,Dor region but located $\sim$\,7$\arcmin$ (100\,pc) from R\,136. We identify
N\,157B\,IRS1 as a candidate protostar from the analysis of its SED and from the
tentative detection of water ice. We have detected no maser emission towards 
N\,157B\,IRS1. Despite the 0.9$\arcmin$ mismatch in position, our experience 
with the 30\,Dor mapping (see for instance analysis of 0539-691C) indicates that
we would have detected emission from this source if stronger than $\sim$1\,Jy. 
We cannot exclude the presence of weaker emission, however water masers are 
normally associated with the earlier (hot core) stage when such objects are 
very weak in the IR \citep{debuizer05} --- IRS1 is the brightest source in our 
sample at near and- mid-IR wavelengths. N\,157B has been identified with 
SNR\,0538-69.1, although its nature as a supernova remnant has been questioned 
\citep{chu04}. The age of the alleged supernova remnant is only $\sim 5000$\,yr 
\citep{wang98} and it is therefore impossible to already have led to the 
formation of stars in the nearby molecular cloud. It is possible that the 
formation of N\,157B\,IRS1 was triggered by feedback from the nearby OB 
association LH\,99, but we have no direct evidence for this. 

\subsection{Triggered star formation in N\,113}

N\,113 is a smaller and perhaps simpler H\,{\sc ii} region than the 30\,Dor
mini-starburst complex. Fig.\,\ref{n113} (bottom right) shows in great detail
the interplay between the ionised gas and the molecular material in N\,113. This
image combines the H$\alpha$\ image (blue) with the 8$\mu$m IRAC/Spitzer image 
(red). We use here the pseudo-continuum image (Sect.\,\ref{halpha}): this image 
contains not only stars (generally blue) but also most of the H$\alpha$ 
emission structure; when combined with the IR image the ionised and neutral gas,
as well as the stellar content become apparent. This composite image presents a
striking example of the ionised bubbles created within the part of a molecular 
cloud facing massive early-type stars (Sect.\,\ref{halpha}).

In Fig.\,\ref{n113}, one can also see that the sequence of bright H$\alpha$
knots A, B, E and G is co-spatial with the brightest 8$\mu$m emission.
\citet{wong06} have recently performed aperture synthesis imaging of this dense 
region in N\,113. The 1.3-cm continuum observations reveal a string of 6 bright 
continuum sources, also located in the dense central region and a large 
molecular clump a few arcseconds away. Fig.\,\ref{n113} (bottom left) shows 
the 8$\mu$m Spitzer image of N113, with several sources identified. The 
positions of two water maser sources \citep{lazendic02} are represented by 
circles. The east-most water maser source is the strongest detected in the 
Magellanic Clouds and it seems to be also associated with an OH maser source 
\citep{brooks97} and a relatively weak continuum source \citep{wong06}. The 
other water maser source is associated with the brightest continuum source that 
\citet{wong06} identify as a compact H\,{\sc ii} region. The approximate 
position of the dense molecular clump is also indicated in Fig.\,\ref{n113}; 
\citet{wong06} propose that the compact H\,{\sc ii} region might trigger further
collapse in this dense clump.

\begin{figure}
\includegraphics[width=84mm]{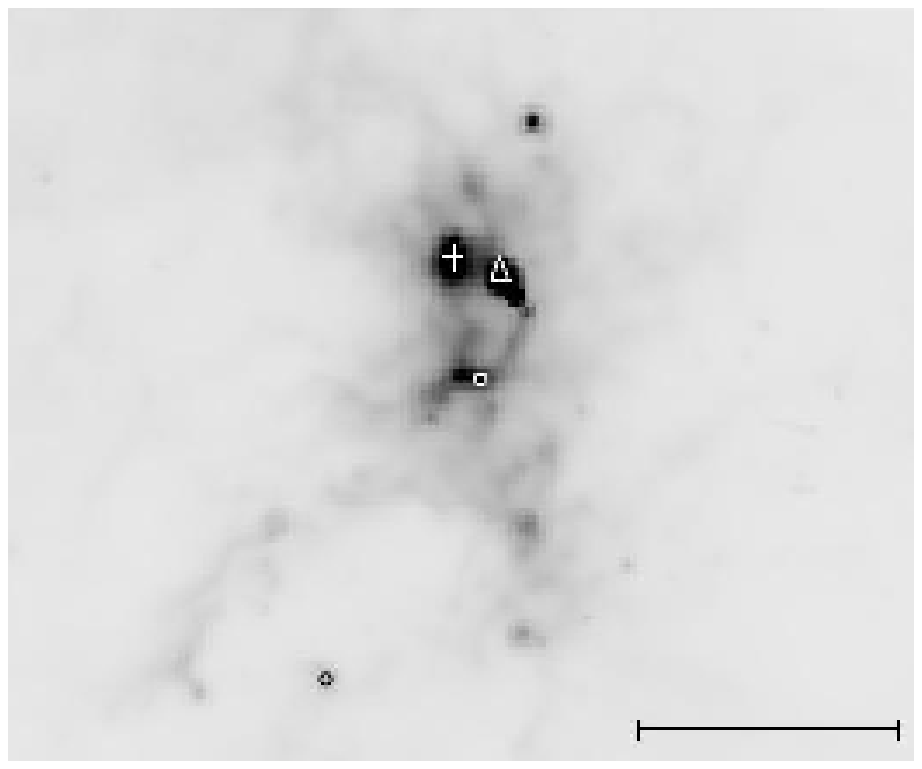}
\includegraphics[width=84mm]{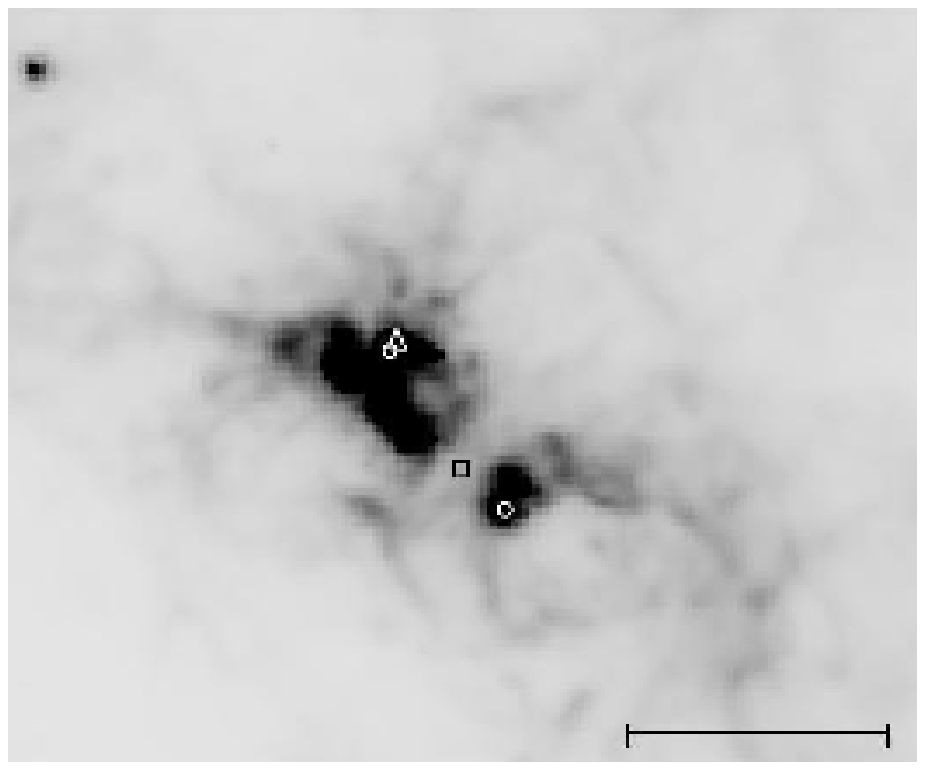}
\caption[]{8\,$\mu$m IRAC/Spitzer images of N\,105A (top) and N\,160A (bottom). 
In each image, the scale indicates 1\,arcmin. Circles represent maser sources 
\citep{sinclair92,lazendic02}, triangles are protostars 
\citep[this work;][]{epchtein84,henning98}, and the square is a molecular core 
(\citealt{bolatto00}). A cross indicates the position of N\,105A\,blob 
(Sect.\,\ref{n105spec}). In both images North is to the top and East to the 
left and the greyscale is linear between $-$20 and 250\,MJy\,sr$^{-1}$. The 
images are centered at $\sim 05^h09^m53^s,-68\degr53\arcmin27\arcsec$ and at 
$\sim 05^h39^m41^s,-69\degr38\arcmin41\arcsec$ respectively for N\,105A and 
N\,160A.}
\label{mir}
\end{figure}

The average molecular hydrogen density in the molecular cloud associated with
N\,113 is estimated at $n_{\rm cloud}\sim200$\,cm$^{-3}$ \citep{wong06} which, 
for a cloud radius of $R_{\rm cloud}\sim15$\,pc, corresponds to a column density
of $N_{\rm cloud}\sim9\times10^{21}$\,cm$^{-2}$. This just exceeds the threshold
for a diffuse cloud to cool and collapse to form stars, 
$N_{\rm critical}\sim10^{21}$\,cm$^{-2}$ \citep{bergin04}. However, 
\citet{wong06} argue that the cloud must be strongly clumped, which would imply 
that most (geometrically) of the cloud is much less dense --- possibly 
$N_{\rm cloud}<N_{\rm critical}$. The dense molecular clump that was detected 
within this cloud has a radius of $R_{\rm clump}\sim1.6$\,pc and a molecular 
hydrogen density of $n_{\rm clump}\sim10^5$\,cm$^{-3}$ \citep{wong06}; this 
yields a column density of $N_{\rm clump}\sim5\times10^{23}$\,cm$^{-2}$, which 
comfortably exceeds the threshold for star formation. The strongest density 
enhancements are located near the interface with the hot gas; thus it is likely 
that the gas has been compressed as a result of feedback from the nearby OB 
stars.

To summarise, the scenario that emerges in N\,113 is that in the compressed 
dense lane of neutral gas and dust star formation is occurring, pinpointed by 
the maser sources in its earliest stages and by the 1.3-cm continuum emission as
the massive stars evolve. N\,113 therefore presents a clear example of star 
formation triggered by the winds from massive stars. A similar scenario seems 
also likely for N\,160A \citep[Fig.\,\ref{mir}; see also][]{nakajima05}.

\subsection{N\,105A, a less evolved H\,{\sc ii} region?}

The situation seems different for N\,105A. Fig.\,\ref{mir} (top) shows the 
8\,$\mu$m IRAC/Spitzer image of N\,105A, dominated by hot dust and PAH emission,
with the positions of N\,105A\,IRS1 (\citealt{epchtein84}; this work), water and
OH maser \citep{lazendic02,brooks97} and methanol maser \citep{sinclair92} 
indicated. Although projection effects may affect our view, the morphology of 
N\,105A shows little evidence for massive star feedback; no shell structure is 
seen in [O\,{\sc iii}] images \citep{ambrocio98}, in spite of the proximity of 
the OB association LH\,31, identified near the molecular cloud \citep{dopita94}.
Furthermore, there seems to be no link between the water and OH maser source 
\citep{lazendic02,brooks97} and the protostar location and any external trigger.
However, the methanol maser to the south \citep{sinclair92} does appear to lie 
in a dense knot at the rim of a cavity in the molecular cloud. A massive young 
stellar object, N\,105A\,IRS1, is already ionising the molecular cloud and also
shows evidence of outflows. It seems likely that it is only a matter of time 
before N\,105A\,IRS1 starts sculpting the N\,105 complex, influencing current 
star formation occurring in the cloud core as signposted by the maser sources.

The molecular cloud associated with N\,105A is a factor five denser than that 
in N\,113 \citep{chin97}. N\,105A is therefore more likely to collapse and 
fragment without the need for an external trigger. Thus N\,105A seems to be at 
an earlier stage of its evolution, whereas N\,113, N\,160A and in particular 
30\,Dor have seen more generations of stars forming prior to the current epoch 
of star formation.

\section{Summary and conclusions}

We have conducted a survey for water maser emission in 30\,Doradus and a sample
of H\,{\sc ii} regions in the Large Magellanic Cloud, to investigate the 
conditions under which current star formation occurs. The locations of the maser
sources are compared with infrared images from the ESO/VLT (at 3.8\,$\mu$m) and 
the Spitzer Space Telescope, and an H$\alpha$ image in the case of N\,113. We 
also present 3$-$4\,$\mu$m spectroscopy to investigate the nature of two 
protostar candidates, in the N\,157B and N\,105A regions, showing evidence for 
the onset of the stellar feedback process in one of them.

Our water maser survey of 30\,Dor uncovered one new source, 0539$-$691C, with a 
velocity consistent with the systemic velocity of the LMC. All detected water 
masers are located in the densest part of the nebula at the interface between 
neutral and ionised gas. N\,105A\,IRS1 shows strong H recombination line 
emission and its SED and IR colours are consistent with an embedded young 
massive star ionising its immediate surroundings. It also shows evidence for 
outflows. A nearby diffuse IR source, N\,105A\,blob, shows both H recombination 
and PAH emission. We identify N\,157B\,IRS1 as an embedded protostar based on the
analysis of its SED and a tentative detection of the 3.1\,$\mu$m water ice 
feature.

In the well-developed H\,{\sc ii} regions 30\,Dor, N\,113 and N\,160A, no water 
masers have been detected deep within the molecular cloud complexes. They are 
always found at the interfaces between molecular cloud and H\,{\sc ii} region. 
This provides strong evidence that feedback from massive stars triggers 
subsequent star formation. Although in the dense cloud N\,105A star formation 
seems to occur without evidence of massive star feedback, the general conditions
in the LMC seem favourable for sequential star formation as a result of 
feedback. 

The wind speed and mass-loss rate of O and B stars decrease at lower 
metallicity (Mokiem \& de Koter, in preparation), thus massive star feedback 
could be weak in metal-poor environments but we see no evidence for this in the 
LMC (metallicity $\sim$40 per cent solar). This is important as the duration and
intensity of star formation epochs in galaxies may depend on the efficiency of 
the local massive star feedback to trigger further star formation. 

\section*{Acknowledgements}
We would like to thank the staff at Parkes, AAT and ESO Paranal for their 
support. We thank the anonymous referee for useful comments. This publication 
makes use of data products from the Two Micron All Sky Survey, which is a joint 
project of the University of Massachusetts and the Infrared Processing and 
Analysis Center/California Institute of Technology, funded by the National 
Aeronautics and Space Administration (NASA) and the National Science Foundation.
We make use of archival images obtained with the Spitzer Space Telescope, which 
is operated by the Jet Propulsion Laboratory, California Institute of Technology
under a contract with NASA. JMO acknowledges financial support by PPARC.

\appendix

\section{Regions with non-detections at 22\,GHz}
\label{other_masers}

We have observed several other H\,{\sc ii} regions at 22\,GHz, both in the Large
and Small Magellanic Clouds. The positions listed in Table\,\ref{no_det} had 
been previously observed by \citet{whiteoak83} and \citet{scalise82}. We did not
detect maser emission at any of these positions, confirming previous 
non-detections \citep{whiteoak83}. \citet{lazendic02} detected maser emission in
N\,159, with a peak flux density of 3.7\,Jy at 247.5\,km\,s$^{-1}$. No source 
was detected in our spectrum. In the SMC, \citet{scalise82} detected water 
maser emission in S\,7 and S\,9 at the systemic velocity of the cloud 
(120\,km\,s$^{-1}$), with peak intensities of 7.4 and 4.2\,Jy, respectively. We 
did not detect any water maser emission at either of these positions. Water
maser emission is known to be intrinsically variable in galactic star forming 
regions, both in intensity and velocity \citep{panagi93,tofani95} therefore it 
is not unexpected that our survey does not detect all previously identified 
maser sources.

\begin{table}
\caption[]{List of non detections in the LMC and SMC.}
\label{no_det}
\begin{tabular}{lccc}
\hline
cloud complex&RA 2000   &DEC 2000   &rms/channel\\
&($^{h\,\,m\,\,s}$)&($\degr\,\,\arcmin\,\,\arcsec$)&(Jy)\\
\hline
\multicolumn{4}{c}{Large Magellanic Cloud}\\
\hline
MC\,69& 05 36 20.7 & $-$69 12 15 &0.10\\
MC\,89& 05 47 09.6 & $-$69 42 15 &0.10\\
N\,11B& 04 56 51.5 & $-$66 24 25 &0.13\\
N\,132D&05 25 01.6 & $-$69 38 16 &0.13\\
N\,158C&05 39 09.3 & $-$69 30 14 &0.09\\
N\,159& 05 39 57.2 & $-$69 44 33 &0.10\\
N\,44D& 05 23 01.0 & $-$68 02 13 &0.10\\
N\,59A& 05 35 24.5 & $-$67 34 52 &0.13\\
\hline
\multicolumn{4}{c}{Small Magellanic Cloud}\\
\hline
N\,66&  00 59 16.9 & $-$72 09 50 &0.11\\
S\,7&	00 46 38.9 & $-$72 40 50 &0.10\\
S\,9&   00 47 30.8 & $-$73 08 20 &0.10\\
\hline
\end{tabular}
\end{table}

\section{R\,Doradus and W\,Hydrae}
\label{rdor}

\begin{figure}
\includegraphics[width=84mm]{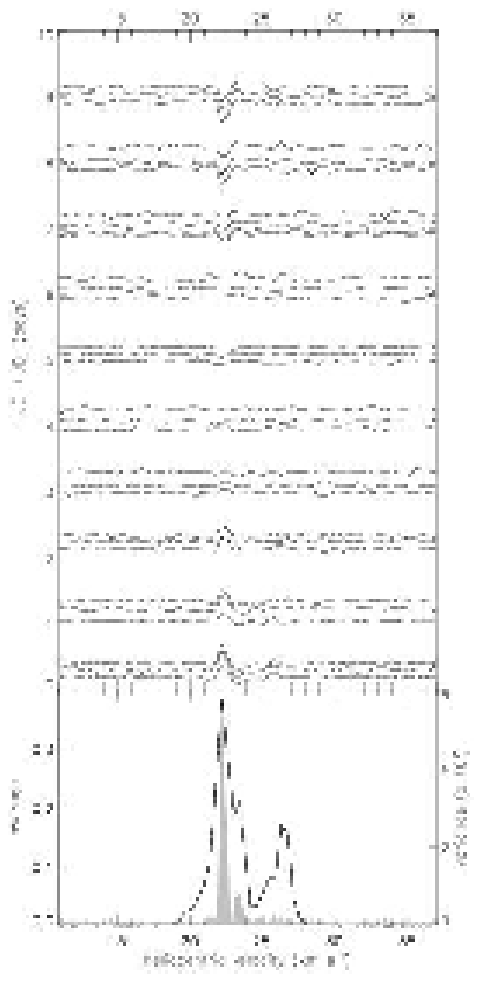}
\caption[]{Time-series of the 22\,GHz observations of R\,Dor. The top panel 
shows each individual spectrum subtracted by the average spectrum. The y-axis
represents the actual time gaps between spectra. The average and variance 
spectra are shown in the panel below. Variability on the maser profiles seems 
to be restricted to the high frequency component at $\sim$\,22\,km\,s$^{-1}$.}
\label{rdor1}
\end{figure}

The nearby galactic red giants R\,Doradus and W\,Hydrae were observed at 
22\,GHz for reference. R\,Dor was observed several times each night. The 
resulting time series of the difference with respect to the average spectrum is
displayed in Fig.\,\ref{rdor1}. The average profile shows two distinct 
components, but only the high-frequency component shows variability. R\,Dor 
pulsates radially in a semi-regular fashion with a period of $P\sim338$\,days, 
but the variability of the water maser emission occurs on a much shorter 
timescale --- 9\,days corresponds to less than 0.03 cycles. 

\begin{figure}
\includegraphics[width=84mm]{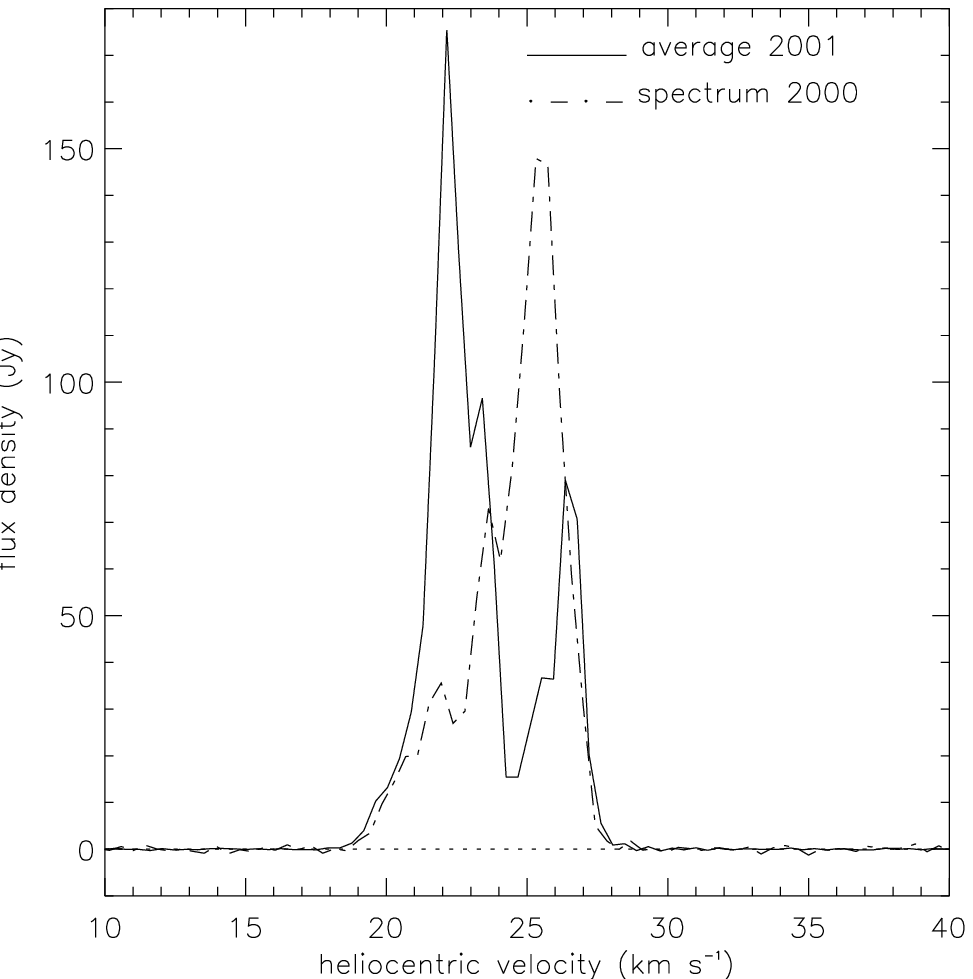}
\caption[]{Average 22\,GHz spectra of R\,Dor, from this work (full line) and
\citet[][ dot-dashed line]{vanloon01b}. In 1.33 pulsation cycles (see text), the
maximum peak intensity has shifted from the low to the high frequency 
component.}
\label{rdor2}
\end{figure}

The average spectrum is compared in Fig.\,\ref{rdor2} to that observed 1.33 
cycles previously \citep[April 2000,][]{vanloon01b}. The 2001 spectrum looks 
more similar to the discovery spectrum \citep*{lepine76}, and is more 
symmetrical around the systemic velocity of the star as derived from the 
centroid of the CO emission profile, $v_\star=24$\,km\,s$^{-1}$ 
\citep{olofsson02}. The double-peaked water maser profile suggests substantial 
radial amplification; this yields an estimate for the outflow velocity in the 
inner part of the dust envelope of $v_{\rm H_2O}\sim2$\,km\,s$^{-1}$ (peaks) to 
4\,km\,s$^{-1}$ (total extent), which is less than the outflow velocity in the 
outer part of the dust envelope as measured in CO, 
$v_{\rm CO}\sim6$\,km\,s$^{-1}$. This implies that the wind is still 
accelerating in the region of the water maser.

\begin{figure}
\includegraphics[width=84mm]{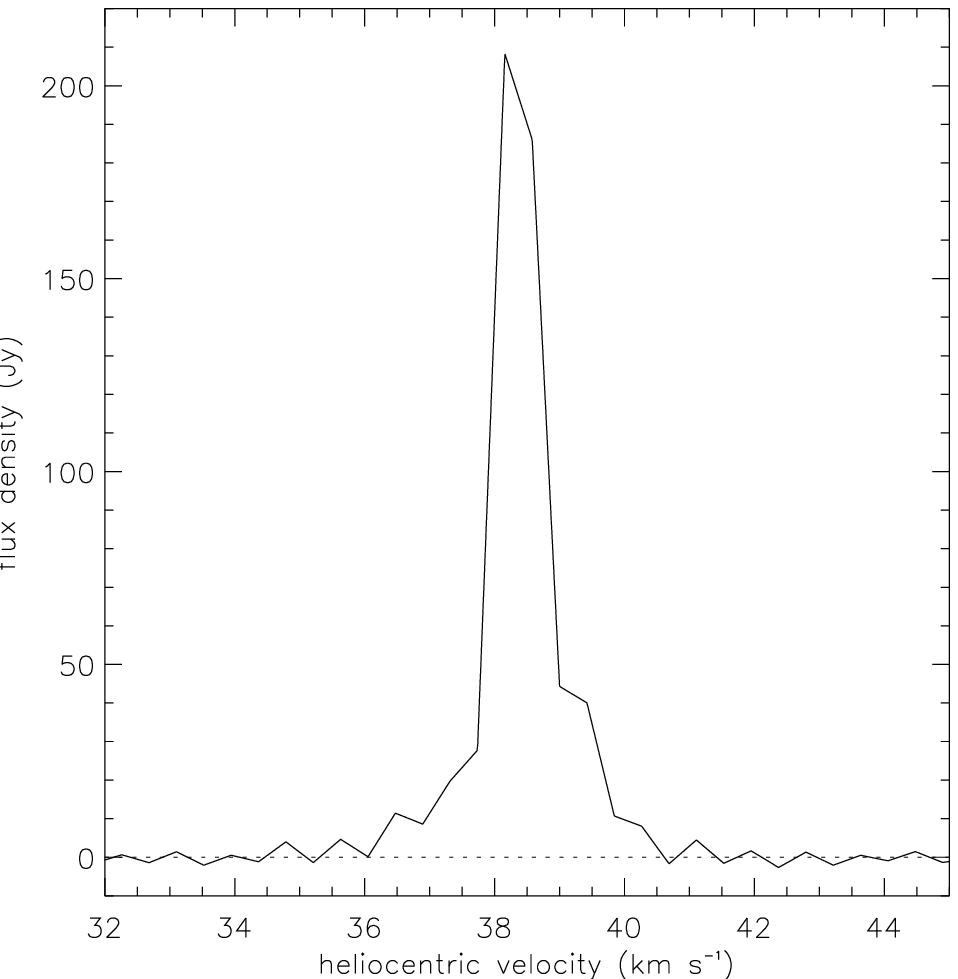}
\caption[]{22\,GHz spectrum of W\,Hya.}
\label{whya}
\end{figure}

The water maser emission profile of W\,Hya (Fig.\,\ref{whya}) has a single peak
near the systemic velocity of the star, $v_\star=39.7$\,km\,s$^{-1}$
\citep{gonzalez98}, suggesting mainly tangential amplification. However, the 
emission profile has a broader pedestal component suggesting radial motions of 
the order of $v_{\rm H_2O}\sim2$\,km\,s$^{-1}$. The water maser is highly 
variable, but the maximum extent of the emission is well constrained and 
suggests $v_{\rm H_2O}<4$ km s$^{-1}$ \citep*{rudnitskii99}. As in the case of 
R\,Dor, this is slower than the outflow velocity in the outer part of the wind 
as measured in CO, $v_{\rm CO}\sim6.5$\,km\,s$^{-1}$ \citep{olofsson02}, and 
consistent with a wind that is accelerated through radiation pressure on dust 
grains.

\bsp

\label{lastpage}

\end{document}